\documentclass[11pt,a4paper]{article}
\pdfoutput=1
\usepackage{jcappub}
\usepackage{graphicx}
\usepackage{subfig}
\usepackage{bm}% bold math

\usepackage{gensymb}

\def\apj{Astrophys.~J.}

\def\prd{Phys.~Rev.~D}
\def\prl{Phys.~Rev.~Lett.}
\def\plb{Phys.~Lett. B.}

\def\cqg{Class.~Quant.~Grav.}
\def\lrl{Living.~Rev.~Relativity.}

\title{Synergy between ground and space based gravitational wave
detectors II: Localisation}

\author[a]{Remya Nair}
\author[a,b]{and Takahiro Tanaka}
\affiliation[a]{Department of Physics, Kyoto
University, 606-8502, Kyoto, Japan}
\affiliation[b]{Yukawa Institute for Theoretical
Physics, Kyoto University, 606-8502, Kyoto, Japan}

\emailAdd{remya@tap.scphys.kyoto-u.ac.jp}
\emailAdd{tanaka@yukawa.kyoto-u.ac.jp}
\abstract
{We study the advantage of combining measurements from future ground and space
based gravitational wave detectors in estimating the parameters of a
black-hole binary coalescence. This is an extension of our previous work (PTEP 053E01 (2016)) where we used pattern averaged
waveform to study non-spinning binaries. In this work we study the localisation and binary plane orientation, including the (non-precessing) spin of binaries. We focus on the third generation terrestrial detector `Einstein telescope' and a proposed space based detector `Deci-Hertz Interferometer Gravitational wave Observatory' (DECIGO). We consider two possible orbits for DECIGO, a helio-centric orbit and a Sun-synchronous geo-centric orbit. We demonstrate that one can obtain order of magnitude improvement in the localisation from the space-ground combined measurements, even with a precursor-DECIGO mission (B-DECIGO). This is especially important for the future of gravitational wave astronomy as improving the localisation accuracy further improves our chances of identifying the host galaxies of these binary systems.}
\begin{document}

\maketitle

\section{Introduction}

We are in an exciting era of gravitational wave (GW) astronomy. After the multiple
GW detections by the LIGO-VIRGO network \cite{gw_det}, and the successful pathfinder mission of
Laser Interferometer Space Antenna (LISA) \cite{lisa_pf}, we can now look forward to a future
where multiple GW detections by both
the space and ground based interferometers will be a norm. Astronomers rely on
various cosmological observations to probe our Universe, some of which include:
the type Ia supernovae, the baryon acoustic oscillations, the cosmic microwave
background, gravitational lensing, etc. Requiring consistency between these
measurements and combining them helped us converge on what we now know as the
standard model of cosmology. Consistency checks between different observations of
the same physical phenomenon help us identify the systematic effects. On the
other hand, combining measurements aids parameter estimations by removing
degeneracies in the parameter space and reducing the errors on the parameter estimates.
The work we present here is an extension of our earlier work, where we demonstrated
the advantage of combining measurements of ground and space based GW interferometers
in estimating parameters of a compact binary coalescence \cite{synergy1}. 

Coalescing compact binaries which are composed of neutron stars (NS) - NS, NS- black
hole (BH), or BH - BH produce GW signals during their inspiral, merger and ringdown
phases. The merger and ringdown phases of at least five such events have been recorded
by the LIGO-VIRGO GW detector network so far \cite{gw_det}. These ground based detectors are sensitive
in the frequency range from a few tens of Hz to a few 1000 Hz. Till now we have seen BH-BH binary mergers
with total mass ranging from $\sim 20~M_{\odot}$ to $\sim 70~ M_{\odot}$.
There are already plans for third generation detectors like the Einstein Telescope (ET) of the European mission and the Cosmic explorer (CE) \cite{next_genGW}. ET and CE will detect binary
coalescence with a higher signal to noise ratio (SNR) and possibly at lower frequencies than the second
generation detectors like advanced LIGO , advanced VIRGO and KAGRA \cite{refET}.
LISA on the other hand, would aim to observe supermassive BH binaries in the frequency range 
0.1 mHz - 1Hz. Many studies have been performed to estimate the binary parameters (and additionally testing gravity theories) with LISA like detectors \cite{berti,yagi_tanaka}. Recently there has been a lot of interest in the possibility of doing cosmography with LISA \cite{kyutoku_seto, pozzo}. There is also a proposal for a Japanese space mission, Deci-Hertz Interferometer Gravitational Wave Observatory (DECIGO) for observing GW around $f \sim 0.1 - 10$ Hz.  In such a scenario, one can
ask how these detectors can complement each other. GW signals from the coalescing binaries that
have passed beyond the LISA band will sweep through the DECIGO band before entering the
frequency range of the ground based interferometers. Hence DECIGO can act as a follow
up for the LISA mission and as a precursor for ground based detectors. 
DECIGO alone can determine the location of NS-NS sources to about an arcminute \cite{nakamura_decigo}, which can aid the electro-magnetic follow up of these events by ground based detectors. Low frequency space detectors
may also help in confirming those GW signals for which only the ringdown signals are detected by the
ground-based detectors. 

%K. Yagi showed that 
%GW signals from an equal-mass intermediate mass BH binary in NGC 6752 can be detected by both the
%DECIGO path finder and ground-based interferometers. The accumulated signal to noise (SNR) in this case
%reaches $\rho$ = 5 at $t \approx 4 \times 10^2$ seconds before the merger \cite{yagi}. 
In this spirit we studied non-spinning NS-BH compact binaries in \cite{synergy1} and for simplicity we
ignored the information of the location of the source
and its orientation with respect to the detector. Instead
we used pattern averaged waveforms to show that low frequency space based interferometers
like DECIGO can complement the observations of ground based measurements from third generation detectors like ET. In
the present paper we extend our earlier work by incorporating information about the location
and orientation of the source and the effect of spin to the GW phase. We further analyse two configurations for the detector orbit of DECIGO, one which is helio-centric and one which is geo-centric. Here we consider BH-BH binaries and focus on the improvement in the localisation accuracy obtained by combining space and ground measurements. In the absence of an electro-magnetic counterpart, such improvements will help us in identifying the host galaxies of these GW sources. In cases where we can identify host galaxies, these GW measurements will be further useful for cosmological studies. 

The paper is organized as follows. In \S \ref{sec:non-spin} and \S \ref{sec:spin} we briefly outline the expressions used for the GW phase in case of non-spinning binaries and spinning (non-precessing) binaries respectively. In \S \ref{sec:error} we discuss the basics of error estimation and the detector noise curves used in this study, and briefly outline the detector orbits in \S \ref{det_orbit} (more details are available in the appendix). We report the results of our analysis  in \S \ref{result}, and discuss implications of our results and some future directions in \S \ref{sec:impli}.

\section{Non-spinning compact binaries}\label{sec:non-spin}
Within general relativity, the post-Newtonian (PN) formalism is used to model the inspiral part
of the binary evolution and the gravitational waveform. Physical quantities like the conserved energy, flux etc. are written
as expansions in a small parameter $(v/c)$, where $v$ is the characteristic speed of the
binary system and $c$ is the speed of light \cite{luc}.  Corrections  of {\it O}$((v/c)^n)$ (counting from the
leading order) are referred to as a $(n/2)$PN order terms in the standard convention. For
non-spinning systems, the GW amplitude is known up to 3PN order, whereas the phase and binary dynamics
are known up to 3.5PN and 4PN order respectively (please refer to \cite{luc,luc02,luc04,luc08,
dam14} and references therein). Spin corrections have been calculated to 2.5 PN order in phase
and 2 PN order in amplitude in \cite{arun09} and Marsat et al., calculated 3 PN and 3.5 PN order
spin-orbit phase corrections \cite{marsat13}.
%in the limit of large 
%signal to noise ratio (SNR), the probability that the GW signal $s(t)$ is characterized by some
%given values of the source parameters $\theta_{\alpha}$ is given by

In an extension to our previous work \cite{synergy1}, we now consider error estimations without averaging
over the relative orientation of the binaries with respect to the detectors, to focus on the sky localisation. 
First we introduce two Cartesian reference frames following \cite{cutler_98}. One is a barred barycentric frame $(\bar{x},\bar{y},\bar{z})$ tied to the ecliptic and centered in the solar system barycentre. In this frame $\bar{\bm{e}}_z$ is normal to the ecliptic and $\bar{x}\bar{y}$-plane is aligned with the ecliptic plane. The second frame (unbarred) is the detector frame $(x,y,z)$ and is centered in the barycentre of the detector. In this frame the direction of the $z$ axis, $\bm{e}_z$, is normal to the detector plane (see Fig. 1 in \cite{YT}).

We assume that the detector output of DECIGO consists of two independent interferometer outputs, much like LISA, and we
introduce the standard mass variables to write the waveform. ${\cal M} = M\nu^{3/5}$ is the chirp mass,
written in terms of the total mass $M=m_1+m_2$ and the symmetric mass ratio $\nu = (m_1 m_2)/M^2$ , where $m_1$ and $m_2$ are the component masses
of the two compact objects in the binary. We begin by writing the frequency domain GW signal $h(f)$ under the stationary phase approximation, for non-spinning compact binaries which are in quasi-circular orbits. Here we use the \emph{restricted PN waveforms},
which keeps the higher order terms in phase but only takes the leading order terms for the amplitude
\cite{cutler}. This simplification is valid in the non-spinning/aligned-spin case, but for binaries with misaligned spins,
the amplitude modulation on precession time scale may also be important to determine the spin parameters
\cite{veccio}. Under the stationary phase approximation, the Fourier component of the
waveform $h_{\alpha}(f)$ ($\alpha=1,2$ is the detector index) can be written as \cite{non_spin_PN}
\begin{equation}
h_{\alpha}(f)= C {\cal A} f^{-7/6}  e^{i\Psi(f)}  \left\lbrace \frac{5}{4} A_{\alpha}(t(f))\right\rbrace
e^{-i\left(\varphi_{p,\alpha}(t(f))+\varphi_D(t(f)) \right)},
\label{waveform}
\end{equation}
where $C=\sqrt{3}/2$ and 1 for the case of DECIGO and ET, respectively. The amplitude ${\cal A}$ and the phase $\Psi(f)$ are given by
\begin{equation}
{\cal A} = \frac{1}{\sqrt{30} \pi^{2/3}} \frac{{\cal M}^{5/6}}{D_L},
\end{equation}
\begin{align}
\Psi(f) &= 2 \pi f t_c - \phi_c +  \frac{3}{128} (\pi {\cal M}f)^{-5/3} \left\lbrace 1+ \left(\frac{3715}{756}+\frac{55}{9}  \nu\right) \nu^{-2/5} (\pi {\cal M} f)^{2/3} - 16\pi \nu^{-3/5}(\pi {\cal M} f)  \right. \nonumber \\
&+\left.  \left(\frac{15293365}{508032}+ \frac{27145}{504} \nu+\frac{3085}{72}\nu^2\right)\nu^{-4/5}(\pi {\cal M} f)^{4/3}+ \pi\left(\frac{38645}{756} -\frac{65\nu}{9}\right) \right. \nonumber \\ & \times \left. \left(1+\log(6^{3/2}\nu^{-3/5}(\pi {\cal M} f))\right) \nu^{-1}(\pi {\cal M} f)^{5/3}
+ \left(\frac{11583231236531}{4694215680} -\frac{640}{3} \pi^2 - \frac{6848}{21}\gamma_E \right. \right. \nonumber\\ 
&+ \left. \left. \left[-\frac{15737765635}{3048192} + \frac{2255}{12} \pi^2\right]\nu+ \frac{76055}{1728}\nu^2 - \frac{127825}{1296} \nu^3 - \frac{6848}{63} \log(64 \nu^{-3/5}(\pi {\cal M} f)) \right) \right. \nonumber \\  
& \times \left. \nu^{-6/5}(\pi {\cal M} f)^2 +  \pi \left(\frac{77096675}{254016} + \frac{378515}{1512} \nu - \frac{74045}{756}\nu^2\right)\nu^{-7/5}(\pi {\cal M} f)^{7/3} \right\rbrace  ,
\end{align}
and the time evolution of the GW is given by
\begin{align}
t(f) &= t_c - \frac{5}{256} {\cal M} (\pi {\cal M}f)^{-8/3} \left\lbrace1+\frac{4}{3} \left(\frac{743}{336}+\frac{11}{4} \nu\right)\nu^{-2/5} (\pi {\cal M} f) ^{2/3} - \frac{32\pi}{5}\nu^{-1} (\pi {\cal M} f) \right. \nonumber \\ 
&+\left.  2\left(\frac{3058673}{1016064} + \frac{5429}{1008} \nu+\frac{617}{144}\nu^2\right) \nu^{-4/5}(\pi {\cal M} f)^{4/3} + \left(\frac{13\pi}{3} \nu - \frac{7729\pi}{252}\right)\nu^{-1}(\pi {\cal M} f)^{5/3} \right. \nonumber \\ 
&+ \left. 15\left(-\frac{10817850546611}{93884313600} + \left[\frac{15335597827}{60963840}+ \frac{1223992}{27720} - \pi^2 \frac{451}{48} -\frac{1041128}{27720}\right]\nu -  \frac{15211}{6912}\nu^2 \right.\right. \nonumber \\ 
&+ \left. \left.  \frac{25565}{5184}\nu^3 + \frac{1712}{105} \gamma_E + \frac{32}{3}\pi^2 + \frac{3424}{1575}\log\left(32768 \nu^{-2/5}(\pi {\cal M} f) ^{2/3}\right) \right) \nu^{-6/5}  (\pi {\cal M} f)^2 \right. \nonumber \\  
&+ \left.  \left(\frac{14809\pi}{378} \nu^2 -\frac{75703\pi}{756}  \nu -\frac{15419335\pi}{127008} \right) \nu^{-7/5} (\pi {\cal M} f)^{7/3}  \right\rbrace.
\end{align}
$D_L$ is the luminosity distance to the binary, $\gamma_E=0.577216 \cdots$ is the Euler's constant, $t_c$ and $\phi_c$ are the time and phase at coalescence, respectively. The waveform polarization phase $\varphi_{p,\alpha}(t)$ and the polarisation amplitude $A_{\alpha}(t)$ are defined as:
%\begin{widetext}
\begin{eqnarray}
A_{\alpha}(t)&=&\sqrt{(1+(\hat{\bm{L}}\cdot\hat{\bm{N}})^2)^2F_{\alpha}^{+}(t)^2+4(\hat{\bm{L}}\cdot\hat{\bm{N}})^2F_{\alpha}^{\times}(t)^2}, \label{Apol} \\
 \cos(\varphi_{\mathrm{p},\alpha}(t))&=&\frac{(1+(\hat{\bm{L}}\cdot\hat{\bm{N}})^2)F^{+}_{\alpha}(t)}{A_{\alpha}(t)}, \label{phipol1} \\
 \sin(\varphi_{\mathrm{p},\alpha}(t))&=&\frac{2(\hat{\bm{L}}\cdot\hat{\bm{N}}) F^{\times}_{\alpha}(t)}{A_{\alpha}(t)}, \label{phipol2}
\end{eqnarray}
%\end{widetext}
where $\hat{\bm{L}}$ is the unit vector parallel to the orbital angular momentum and $\hat{\bm{N}}$ is the unit vector pointing toward the centre of mass of the binary system. $F_{\alpha}^+$ and $F_{\alpha}^{\times}$ are the beam pattern functions for the plus and cross polarisation modes for the detectors (see Appendix \ref{app_detResponse}). $(\bar{\theta}_{\mathrm{S}},\bar{\phi}_{\mathrm{S}})$ represents the direction of the source in the barred-barycentric frame. We discuss the frequency cut-offs and the detector orbits in sections \S \ref{sec:error} and \S \ref{det_orbit} respectively.
\section{Spinning compact binaries}\label{sec:spin}
The efficiency of detection of GW signals from inspiraling binaries and the accuracy of the parameter estimation depend crucially on the accuracy of the templates used for matched filtering. Hence, while constructing these templates it is important to consider all the physical parameters which may effect the GW signal. The non-spinning or small spin approximation may work in some systems but it is important to consider the effect of spin on the waveform. In this work we restrict ourselves to spin-aligned (or antialigned), nonprecessing BH-binary systems. There are studies that show that including precession breaks the degeneracies in the parameter space and improves parameter estimation \cite{Lang06,Vecc04,Lang11}. We will consider this effect in a future publication. 

We now give the expression for the gravitational waveform used for the spin-aligned (anti-aligned) systems. Again we only take the post-Newtonian corrections in phase. Here we include spin corrections to the phase which include spin-orbit corrections at 1.5PN, 2.5PN, 3PN and 3.5PN order, and spin-spin corrections at 2PN order \cite{wade,arun09,marsat13}.

\begin{align}
\Psi^{\rm spin}(f) &= 2 \pi f t_c - \phi_c + \frac{3}{256} (\pi {\cal M}f)^{-5/3} \left\{1+\left(\frac{3715}{756} + \frac{55}{9}\nu\right) \nu^{-2/5} (\pi {\cal M}f)^{2/3} + \left(4  \beta - 16 \pi\right) \nu^{-3/5}
 \right. \nonumber \\ 
\nonumber
&\times \left. (\pi {\cal M}f)+  \left(\frac{15293365}{508032} +  \frac{27145}{504}\nu + \frac{3085}{72}\nu^2 - 10  \sigma\right) \nu^{-4/5} (\pi {\cal M}f)^{4/3} 
 + \left(\frac{38645\pi}{756} - \frac{65\pi}{9}\nu -  \gamma\right) \right. \\ 
 \nonumber
 &\times \left. \left(1+3 \ln\left( \nu^{-1/5} (\pi {\cal M}f) \right)\right) \nu^{-1} (\pi {\cal M}f)^{5/3} + \left[\frac{11583231236531}{4694215680} - \frac{6848}{21}  \gamma_{\rm E} - \frac{640 \pi^2}{3} \right. \right. \\
 \nonumber
 &+ \left. \left. \left(\frac{2255 \pi^2}{12} - \frac{15737765635}{3048192}\right) \nu \right . + \frac{76055}{1728} \nu^2 - \frac{127825}{1296} \nu^3\left . - \frac{6848}{21} \ln\left(4 v_k\right) + \left(160 \pi \beta - 20 \xi\right) \right] \right. \\
 \nonumber
 &\times \left. \nu^{-6/5} (\pi {\cal M}f)^{2}   \left[ \frac{77096675\pi}{254016} + \frac{378515\pi\nu }{1512} - \frac{74045\pi\nu^2}{756} + \alpha \left(-20 \zeta + \gamma \left(-\frac{2229}{112} - \frac{99\nu}{4}\right) \right.  \right . \right .\\
  \nonumber
&+\left. \left . \left .  \beta \left(\frac{43939885}{254016}+\frac{259205\nu}{504} + \frac{10165 \nu^2}{36} \right)\right) \right] \nu^{-7/5} (\pi {\cal M}f)^{7/3} \right \} \ ,
\label{spinphase}
\end{align}
with
\begin{eqnarray}
\beta &=& \sum_{i=1}^2\left(\frac{113}{12} \left(\frac{m_i}{M}\right)^2 + \frac{25}{4}\nu\right) \vec \chi_i \cdot \hat{\bm{L}}\ , \nonumber \\
\sigma &=& \nu \left[\frac{721}{48} \left(\vec \chi_1 \cdot \hat{\bm{L}}\right) \left(\vec \chi_2 \cdot \hat{\bm{L}} \right) - \frac{247}{48}\left( \vec \chi_1 \cdot \vec \chi_2\right) \right ] \sum_{i=1}^2 \left \{\frac{5}{2}  \left(\frac{m_i}{M}\right)^2 \left[3 \left(\vec \chi_i \cdot \hat{\bm{L}} \right)^2 - \chi_i^2\right] \right . \nonumber \\
&&\left .+ \frac{1}{96} \left(\frac{m_i}{M}\right)^2\left[7  \chi_i^2 - \left( \vec \chi_i \cdot \hat{\bm{L}} \right)^2\right]\right \} \ , \nonumber \\
\gamma &=& \sum_{i=1}^2 \left [ \left ( \frac{732985}{2268} + \frac{140}{9} \nu\right)\left(\frac{m_i}{M}\right)^2 \right .\left. + \nu \left(\frac{13915}{84} - \frac{10}{3} \nu\right)\right]\vec \chi_i \cdot \hat{\bm{L}} \ , \nonumber \\
\xi &=& \sum_{\i=1}^2 \left[ \frac{75 \pi}{2} \left(\frac{m_i}{M}\right)^2 + \frac{151 \pi}{6} \nu\right] \vec \chi_i \cdot \hat{\bm{L}}  \ , \nonumber \\
\zeta &=& \sum_{i=1}^2 \left[ \left(\frac{m_i}{M}\right)^2 \left (\frac{130325}{756} - \frac{796069}{2016} \nu + \frac{100019}{864} \nu^2 \right)  + \nu \left(\frac{1195759}{18144} - \frac{257023}{1008} \nu  \right. \right. \nonumber \\
&& \left. \left. + \frac{2903}{32} \nu^2 \right) \right ] \vec \chi_i \cdot \hat{\bm{L}},
\end{eqnarray}
where $\sigma$ is a spin-spin correction and $\beta$, $\gamma$, $\xi$ and $\zeta$ are spin-orbit corrections. $\alpha$ is either 1 or 0 to turn on or off the 3PN and 3.5PN order spin corrections to the phase.
Here, $\vec \chi_i = \vec S_i / m_i^2$ are the dimensionless spins of the $i$th compact object of the binary. For BHs, the dimensionless spin parameters $\vec \chi_i$ is smaller than unity, while for NS, they can be larger in principle but are thought to be typically much smaller than unity. One can decompose the component spins $\chi_i$ into a symmetric and an antisymmetric combination,
\begin{eqnarray}
\vec \chi_s &=& \frac{1}{2} \left (\vec \chi_1 +\vec  \chi_2\right),  \\
\label{chi}
\vec \chi_a &=& \frac{1}{2} \left( \vec  \chi_1 - \vec \chi_2 \right) \ .
\end{eqnarray}
Here, $\vec \chi_a \cdot \hat{\bm{L}} = \pm |\vec \chi_a|$ and $\vec \chi_s \cdot \hat{\bm{L}} = \pm |\vec \chi_s|$.  Hence in addition to ${\cal M},\nu,t_c,\phi_c$, and the four angles specifying the location and orientation of the binary with respect to the detector, we now also have spin correction parameters.

In the following section we introduce the Fisher matrix approach we use for error estimation, and provide the noise curves used in our analysis. 
\section{Error estimation}\label{sec:error}
GW signals coming from the inspiral of compact binaries are very weak. To look for these signals in the noisy output of the GW interferometers, the technique of matched filtering is used \cite{bAllen}. Waveforms in a template bank are fitted to the detector output to extract signals that may match them. As can be expected, the
effectiveness of such a method relies on accurate modeling of the GW signals as incorrect modeling can 
lead to systematic errors in the parameter estimations or missing the signal altogether.
Next, to estimate statistical errors in the parameter estimates, the standard Fisher Matrix method can be used. We briefly give an overview
of this method in this section but please refer to \cite{cutler,finn,valli08} for excellent reviews and details ({\it and} shortcomings)
of the method. 

We start by assuming that the GW signal depends on the parameter vector $\boldsymbol{\theta}$. 
So in the non-spinning case $\boldsymbol{\theta}=\lbrace \log{\cal M}, \nu, t_c, \phi_c, \bar{\theta}_{\mathrm{L}}, \bar{\theta}_{\mathrm{S}}, \bar{\phi}_{\rm{L}}, \bar{\phi}_{\mathrm{S}} \rbrace$. For the spinning case we find that only the coefficient at the leading order spin corrections at 1.5PN and 2PN are enough to specify the higher order terms, assuming aligned spins. Hence we focus only on the leading order corrections here and we have $\boldsymbol{\theta}=\lbrace \log{\cal M}, \nu, t_c, \phi_c, \bar{\theta}_{\mathrm{L}}, \bar{\theta}_{\mathrm{S}}, \bar{\phi}_{\rm{L}}, \bar{\phi}_{\mathrm{S}}, \beta,\sigma \rbrace$. We study BH-BH binaries with component masses 30 $M_{\odot} +$ 40 $M_{\odot}$ located at a distance of 3 Gpc. The fiducial values of other parameters are $t_c=\phi_c=\beta=\sigma=0$. Choices for angles $\bar{\theta}_{\mathrm{L}}, \bar{\theta}_{\mathrm{S}}, \bar{\phi}_{\rm{L}}, {\rm and} ~\bar{\phi}_{\mathrm{S}}$ are explained in \S \ref{nonS}.

Now we write the standard expressions used for obtaining the Fisher matrix. The noise weighted inner product
of two waveforms $h_1(t)$ and $h_2(t)$ is defined as
\begin{equation}
(h_1,h_2)=2 \int _0 ^{\infty} \frac{\tilde{h}_1^*(f) \tilde{h}_2(f)+
\tilde{h}_2^*(f) \tilde{h}_1(f)}{S_n(f)}df.
\label{in_pro}
\end{equation}
$\tilde{h}_1(f)$ and $\tilde{h}_2(f)$ are the Fourier transforms of $h_1(t)$ and $h_2(t)$, respectively, and ``$*$'' represents 
the complex conjugation. To account for the frequency dependent sensitivity of the GW interferometers, the outputs are
weighted by the power spectral density of detector noise $S_n(f)$. 
The Fisher matrix is defined as \cite{cutler}
\begin{equation}
\Gamma_{ij}\equiv \left(\frac{\partial { h}}{\partial \theta_i},
\frac{\partial { h}}{\partial \theta_j}\right).
\label{fishm}
\end{equation}
In the limit of large SNR, the probability that the signal is characterized by the chosen parameters $\boldsymbol{\theta}$ is given by
\begin{equation}
P(\Delta\theta^i)\propto e^{-\Gamma_{ij}\Delta\theta^i\Delta\theta^j /2}.
\end{equation}
In the limit of large SNR, and stationary Gaussian noise, the inverse of the Fisher matrix gives the error covariance matrix $\Sigma$ of the parameters.  The diagonal elements
of this covariance matrix give the root mean square error in the estimate of the parameters:
\begin{equation}
\sqrt{\left\langle(\Delta\theta^i)^2\right\rangle}=\sqrt{\Sigma^{ii}}.
\end{equation}
The angular resolution $\Delta \Omega$ is defined as 
\begin{equation}
\Delta \Omega \equiv 2 \pi |\sin \bar{\theta}_{\mathrm{S}}| \sqrt{\Sigma_{\bar{\theta}_{\mathrm{S}},\bar{\theta}_{\mathrm{S}}} \Sigma_{ \bar{\phi}_{\mathrm{S}}, \bar{\phi}_{\mathrm{S}}} - \Sigma^2_{ \bar{\theta}_{\mathrm{S}},\bar{\phi}_{\mathrm{S}}}}.
\end{equation}
Note that the Fisher matrix formalism is limited to high SNR cases (see \cite{cornish06,rodri13,valli11,cho13,cho14} for case studies where the Fisher matrix formalism fails). In spite of its limitations, the Fisher matrix method
is the simplest and one of the most inexpensive ways to infer parameter uncertainties from future surveys. In this work our main aim is to study the synergy (or lack thereof) between ground-space detectors in a qualitative way and Fisher matrix method is accurate enough for this purpose. Moreover, we only study high SNR cases, SNR$>650$ for DECIGO and SNR$>20$ for B-DECIGO and the use of Fisher matrix for error estimation is justified. Hence we will adopt the Fisher matrix method for our error estimations. 

It is fairly straightforward to extend the Fisher matrix method to joint measurements. One merely needs to add the Fisher matrices of
the individual measurements, $\Gamma_{\rm Combined} = \Gamma_{1}+\Gamma_{2}$,
and then invert the summed matrix. This is how we obtain combined estimates while analyzing the synergy effect
between DECIGO and ET. The covariance matrix for the combined measurement and the corresponding
error estimate is given as
\begin{eqnarray}
\Sigma_{\rm Combined} &=& \Gamma_{\rm Combined}^{-1} ~,\\
\Delta \theta_{\rm Combined}^{i} &=& \sqrt{\Sigma_{\rm Combined}^{ii}} ~.
\label{Ccomb}
\end{eqnarray}
Combining measurements may help in resolving the degeneracy between parameters and hence improve the parameter estimates.
%------------------------------------------------
\subsection{Noise curves}\label{secnoise}
The output of a GW interferometer $s(t)$, is composed of two components:
the signal $h(t)$ and the detector noise $n(t)$, $s(t)=h(t)+n(t)$. 
We will assume that the detector noise is stationary and Gaussian, with zero mean $\left\langle \tilde{n} \right\rangle =0$ (note that this is not the case in actual observations). Here angular brackets denote average over different noise realizations. The assumption of stationarity ensures
that the different Fourier components of the noise are uncorrelated. The (one-sided) noise power spectral density $S_n(f)$ 
is then given by
\begin{equation*}
\left\langle \tilde{n}(f)\tilde{n}(f') \right\rangle =
\frac{1}{2}\delta(f-f')S_n(f).
\end{equation*}
The square root of the power spectral density is commonly used to describe the sensitivity of a GW interferometer. When $S_n(f)$  is integrated over positive frequencies, it gives mean square amplitude of the noise in the detector~\cite{moore}. Below we give the expressions of the  noise spectral densities of DECIGO and ET used in this work.
\subsubsection*{DECIGO}
The Decihertz Interferometer Gravitational Wave Observatory (DECIGO) is a future plan of a space
mission initially proposed by Seto et al. \cite{Seto:2001qf}, with an aim of detecting GWs in the
frequency range $f \sim 0.1-10$ Hz. Owing to its sensitivity range, DECIGO
would be able to observe inspiral sources that have advanced beyond the frequency band
of space based detector like LISA, but which have not yet entered the ground detector
band. The following form for the DECIGO noise curve is adopted from Yagi and Seto \cite{seto}:
\begin{equation}
\begin{split}
S_n(f)=7.05 \times 10^{-48} \left[ 1+ \left(\frac{f}{f_p}\right)^2 \right]
+ 4.8 \times 10^{-51} \left(\frac{f}{1 \mbox{Hz}}\right)^{-4} \frac{1}
{1+\left(\frac{f}{f_p}\right)^2}\\+5.33 \times 10^{-52}\left(\frac{f}{1 \mbox{Hz}}\right)^{-4} 
\mbox{Hz}^{-1},
\label{decigo_noise}
\end{split}
\end{equation}
where $f_p = 7.36$ Hz. 

\subsubsection*{ET}
\begin{figure}[h]
\centering  
{\includegraphics[width=3.4in,height=2.4in]{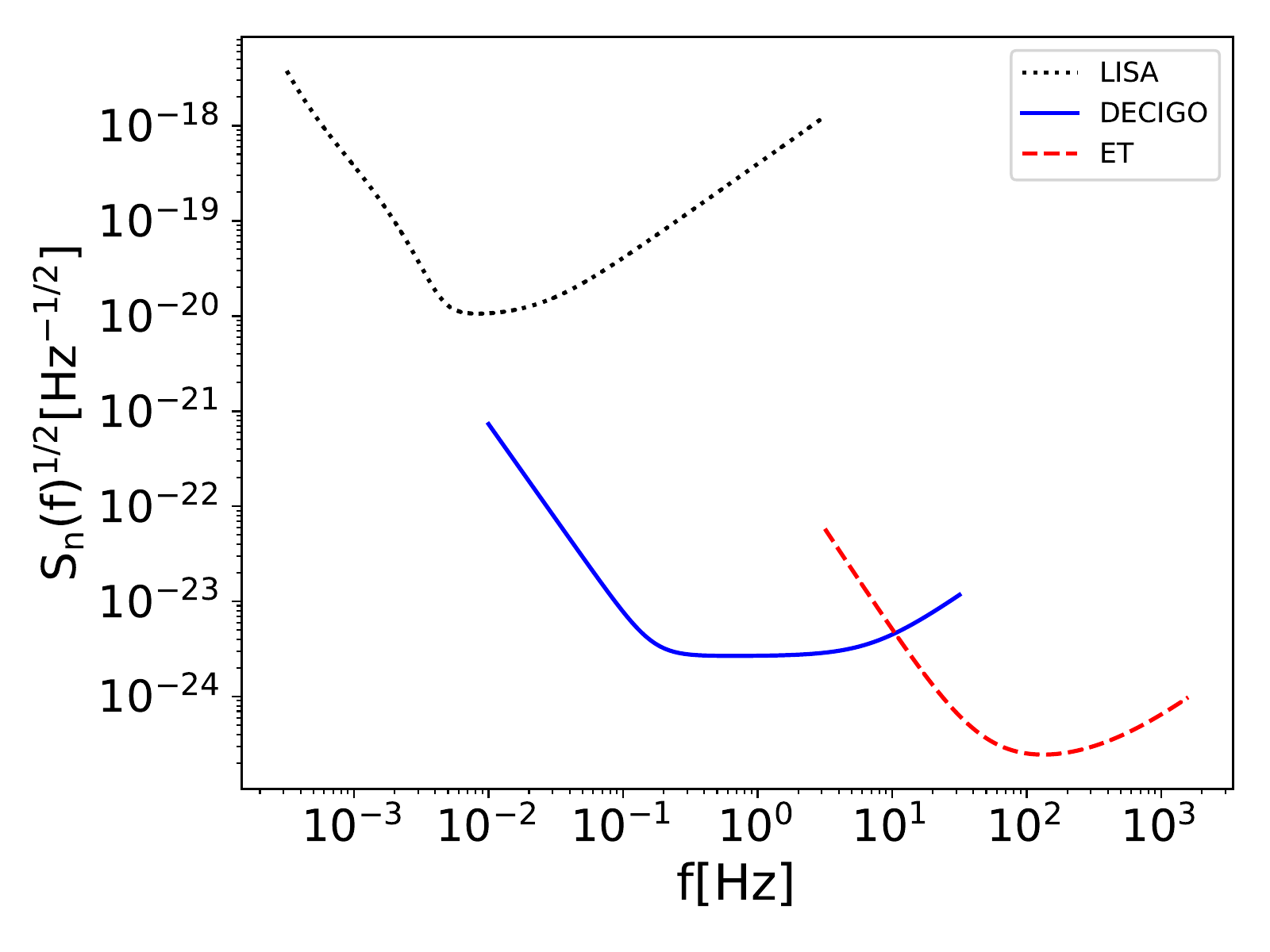}}
\caption{Noise spectra for the different GW detectors. The analytical noise curve for DECIGO, ET and LISA are taken from \cite{seto}, \cite{ajith} and \cite{lisa_noise} respectively.}
\label{nc}
\end{figure}

The Einstein Telescope (ET) is a European commission project. The aim here is to develop a third generation 
GW observatory and achieve high SNR GW events at distances that are comparable with the sight distance
of electromagnetic telescopes. We adopt the noise curve given by Keppel and Ajith \cite{ajith} which was obtained
by assuming ET to be a single L-shaped interferometer with a 90$\degree$ opening angle and arm length of 10 km (ET-B):
\begin{equation}
\begin{split}
S_n(f)=10^{-50} \left[2.39 \times 10^{-27}\left(\frac{f}{f_0}\right)^{-15.64} +0.349 
\left(\frac{f}{f_0}\right)^{-2.145} +
1.76 \left(\frac{f}{f_0}\right)^{-0.12} \right. \\ \left. +0.409 \left(\frac{f}{f_0}\right)^{1.1} 
\right]^2 \mbox{Hz}^{-1},
\end{split}
\end{equation}
where $f_0=100$ Hz.  The noise curves for the GW detectors are plotted in Fig. \ref{nc} for easy reference.
\subsection{Varying design sensitivity for B-DECIGO}
There is a proposal for a precursor mission for DECIGO called B-DECIGO \cite{bdecigo}. The design sensitivity of B-DECIGO is still to be decided conditional on the scientific gain. Thus, in addition to assuming
the noise sensitivity as given in Eq.\eqref{decigo_noise}, we also study the synergy effect between the space detector and ET
by varying the sensitivity of DECIGO. For brevity, we simply call it B-DECIGO. 
The detector sensitivity is changed by scaling the design sensitivity curve \eqref{decigo_noise} uniformly over all frequencies:
\begin{equation}
S_n(f)^{\rm B-DECIGO}={\cal K} S_n(f)^{\rm DECIGO}, 
\label{sdecigo}
\end{equation}
with a constant ${\cal K}$. 
%While reporting the SNRs in the results we only present the SNRs for the original (estimated) design sensitivities. Obtaining the
%SNRs for the scaled-DECIGO is straightforward. Note that the SNR also scales according to 
%the distance to the source. The SNR reported in the tables are just reference values for our choice of fiducial source distance (200 Mpc).
\subsection{Frequency cutoffs for the integral in Fisher Matrix}\label{sec:freq_cut}
Now we discuss our choice of the frequency cutoffs, ($f_{{\rm  in}},f_{{\rm  fin}}$), for the integral in Eq. \eqref{fishm}. First we introduce cutoff frequencies ($f_{{\rm  low}},f_{{\rm  high}}$) for the two GW detectors considered in this work. For DECIGO we choose $f_{{\rm  low}} =0.01$ Hz, $f_{{\rm high }}=20$ Hz and for ET we choose 
$f_{{\rm low }} =10$ Hz, $f_{{\rm high }}=100$ Hz. We assume one year of observation in the space based band and the lower cutoff frequency for the integral is chosen as $f_{{\rm in }}={\rm max}(f_{{\rm year }},f_{{\rm low }})$, where $f_{{\rm year }}$ is the GW frequency 1 year before merger. We choose the upper cutoff frequency of the integral $f_{{\rm fin}}$ as $f_{{\rm fin}}={\rm min}(f_{{\rm high }},f_{{\rm LSO }})$, where $f_{{\rm LSO }}=1/(6^{3/2}\pi M)$ is the approximate frequency corresponding to the last stationary orbit. For the binary system we consider here, $f_{{\rm LSO }}\approx 62.9$ Hz.

There is another proposed space based mission called LISA or eLISA (\cite{lisa_noise}). eLISA design is most sensitive at milli-Hertz frequencies, and hence more suitable for observing inspirals with much larger total mass. The binary systems that can be observed by the ground detectors do not have an extremely large total mass. We do not expect to observe high SNR events from the inspiral phase of such binaries using eLISA. If one considers $\sim$1 year of observation time, the SNR wouldn't be enough to observe these low mass binaries using eLISA (multiband astronomy with eLISA may be possible with 5-10 years of observation, see \cite{elisa_multi}). 

In the following section we give a brief overview of the detector orbits for ET and DECIGO. Specific expressions for detector orbits are provided in appendix \ref{app_detOrbit}.
\section{Detector orbits}\label{det_orbit}
There are three detector orbits we consider here. The terrestrial detector, ET, would rotate, according the the Earth's rotation, once every day and would orbit around the Sun every year. Similarly, if DECIGO is in a geo-centric orbit, it will reorient by orbiting around the Earth every $T_E$ (few hours) and orbit the Sun once a year. On the other hand if DECIGO is in a helio-centric orbit, it will orbit the Sun once per year. Both these orbital motions (helio-centric and geo-centric) contain information about the location of the source. For ET, the reorientation of the detector (with Earth's rotation) has more contribution than its motion around the Sun (which is negligible). We briefly highlight the detector orbits in this section and for the expressions corresponding to the various detector orbits and terms to be used in \eqref{Apol}-\eqref{phipol2}, we refer the readers to  appendix  \ref{app_detOrbit}. 
\subsection*{Orbit for ET}
The detector orbit of ET is obtained by assuming that it will be situated near Sardinia, Italy (lattitude 39$\degree$ N). As mentioned earlier, for such an orbit we account for the re-orientation of the detector with the Earth's rotation and the Doppler effect due to the Earth's motion around the Sun. 

\subsection*{Helio-centric orbit for DECIGO}
We refer the readers to \cite{Seto:2001qf} for the original plan for DECIGO. This orbit is helio-centric much like LISA, but the interferometer arm-lengths are much shorter.

\subsection*{Geo-centric orbit for B-DECIGO}
The orbit of B-DECIGO is not fully determined yet. A possible alternative to the helio-centric orbit is a geo-centric orbit where the detector orbits the Earth in the same way (record plate orbit) as in the case of helio-centric orbit proposed in \cite{Seto:2001qf}. We further assume a Sun-synchronous orbit (which allows the  detector to receive sunlight constantly). The orbital plane will precess because of the spin-orbit coupling and there will be added perturbations to the orbit because of the Moon, but the effect would be negligible. For this test case, we fix the distance of the detector from the Earth's surface at $\sim$2600 km ($T_e \sim2.3$ hours) with the detector plane having an inclination of  $\epsilon \sim 85.6 \degree$ with the ecliptic plane. The precession of the orbital plane of the record plate orbit is neglected. We account for the re-orientation of the detector with the rotation around the Earth in addition to the Doppler effect due to the Earth motion around the Sun. 

In the next section we discuss our findings.  Results obtained when assuming a geo-centric orbit for DECIGO are  labeled with `G', while those from the helio-centric orbit configuration are labeled with `H'.

\section{Results}\label{result}
In this section we present the results of our analysis for the non-spinning and spin-aligned BH-BH binaries. We would like to emphasize that our results should be understood in a qualitative way. Since we only use the information coming from the inspiral phase of the binary coalescence, our results will contradict with studies that use the information from the inspiral-merger-ringdown evolution.  Also, with ET-only measurements, we do not expect any localisation information and it is not possible to put constraints on the parameters. One way to get around this issue is to use a multi-detector terrestrial network composed of ET, LIGO-VIRGO, LIGO-INDIA and KAGRA. We leave this discussion for a future publication. Nonetheless, ET measurements help to remove degeneracy between parameters when combined with DECIGO measurements, thereby improving the error estimates. The error estimates obtained from DECIGO in helio-centric and geo-centric orbit configurations are similar in magnitude but we note that the synergy between the ground and space based measurements is larger for the case of helio-centric orbit, especially for the localisation errors. This is because the distance between the ground and space detector is larger in the case of the helio-centric orbit which aids in the parameter estimation, especially for sky localisation. 
In the following sections we report DECIGO-only and joint DECIGO-ET estimates for the no-spin and aligned-spin cases. 
\subsection{Non-spinning BH-BH binary}\label{nonS}
\begin{figure}[ht]
		\hspace{-0.6cm}
		\subfloat[Part 3][]
		{
			\includegraphics[width=3.1in,height=2.7in]{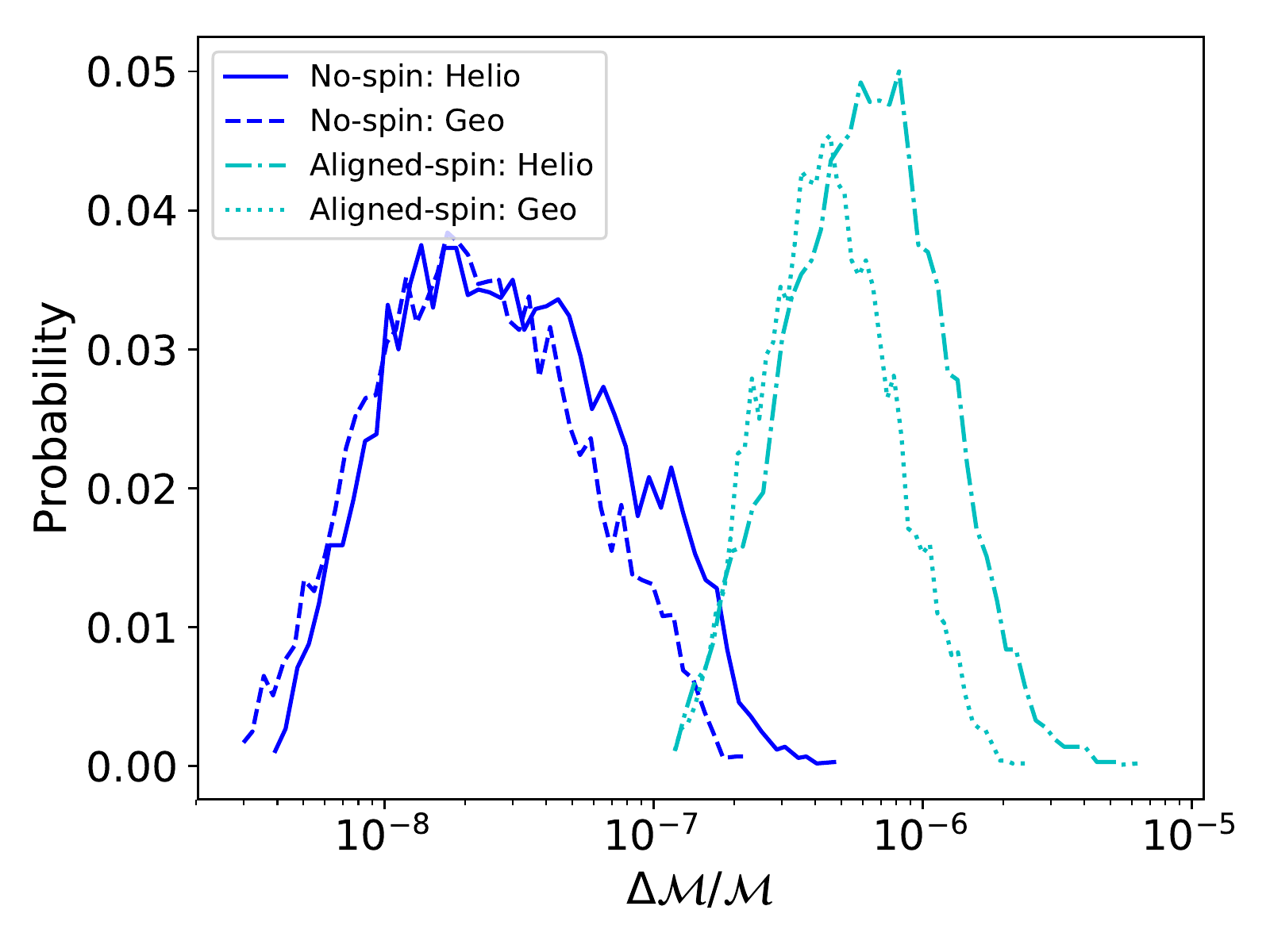}
		}
		\subfloat[Part3][]
		{
			\includegraphics[width=3.1in,height=2.7in]{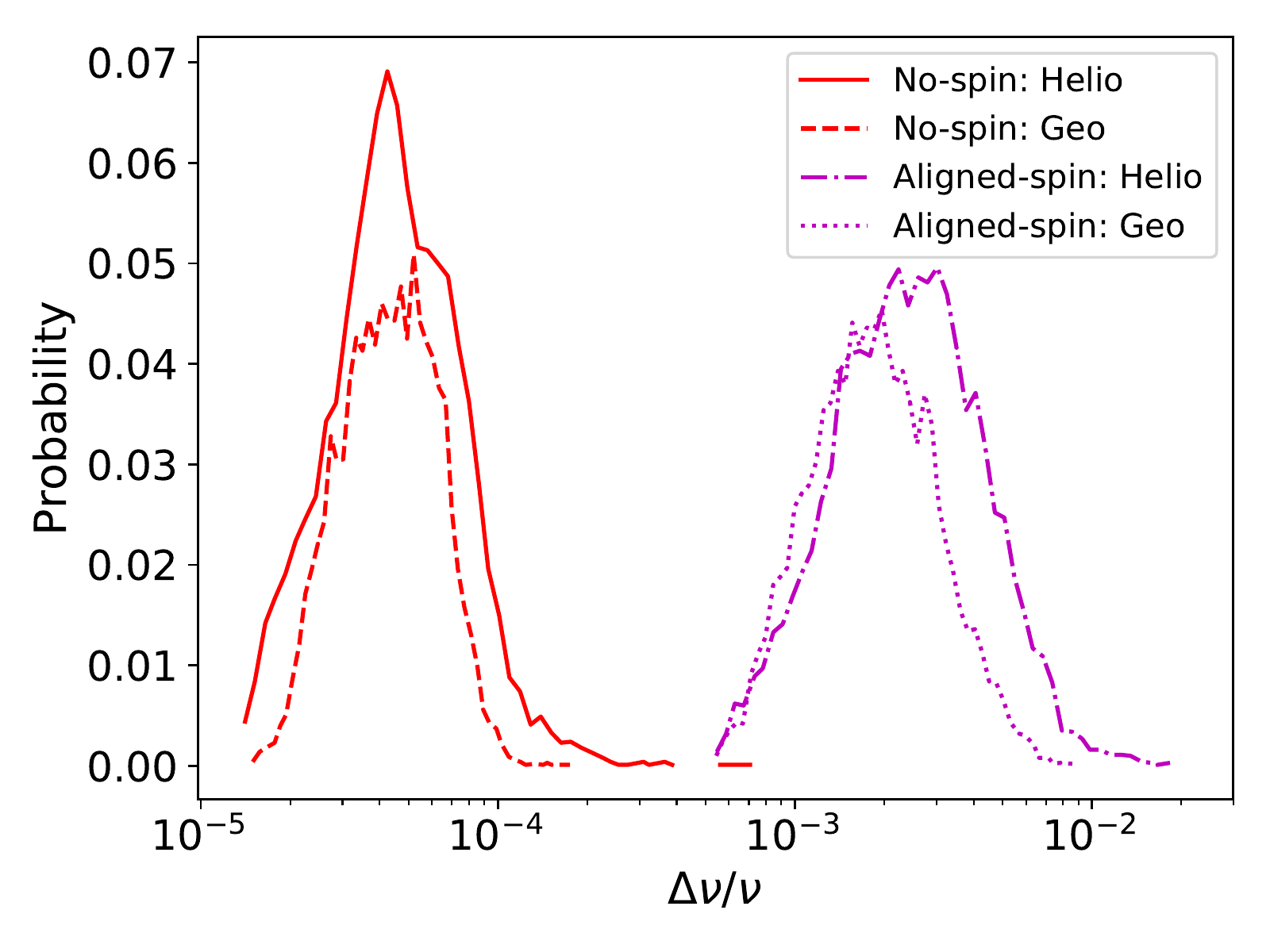}
		}
		\caption{DECIGO: The left panel (a) shows results obtained for the error estimates on the chirp mass ${\cal M}$ for the helio-centric and geo-centric orbit configuration for DECIGO, for the no-spin (blue-solid/dashed curve) and spin-aligned (cyan-dot-dashed/dotted curve) cases. The right panel shows similar results obtained for the  symmetric mass ratio $\nu$ for the no-spin (red solid/dashed curve) and spin-aligned (magenta dot-dashed/dotted curve) cases. We have considered $10^4$ (30$M_{\odot}$+40$M_{\odot}$) BH-BH binaries (at 3 Gpc) distributed uniformly over the sky (see section \ref{nonS}).
		}
		\label{figD_mc_nu}
\end{figure}

\begin{figure}[h]
	\vspace{0.1cm}
	\hspace{-0.6cm}
	\subfloat[Part3][]
	{
		\includegraphics[width=3.1in,height=2.75in]{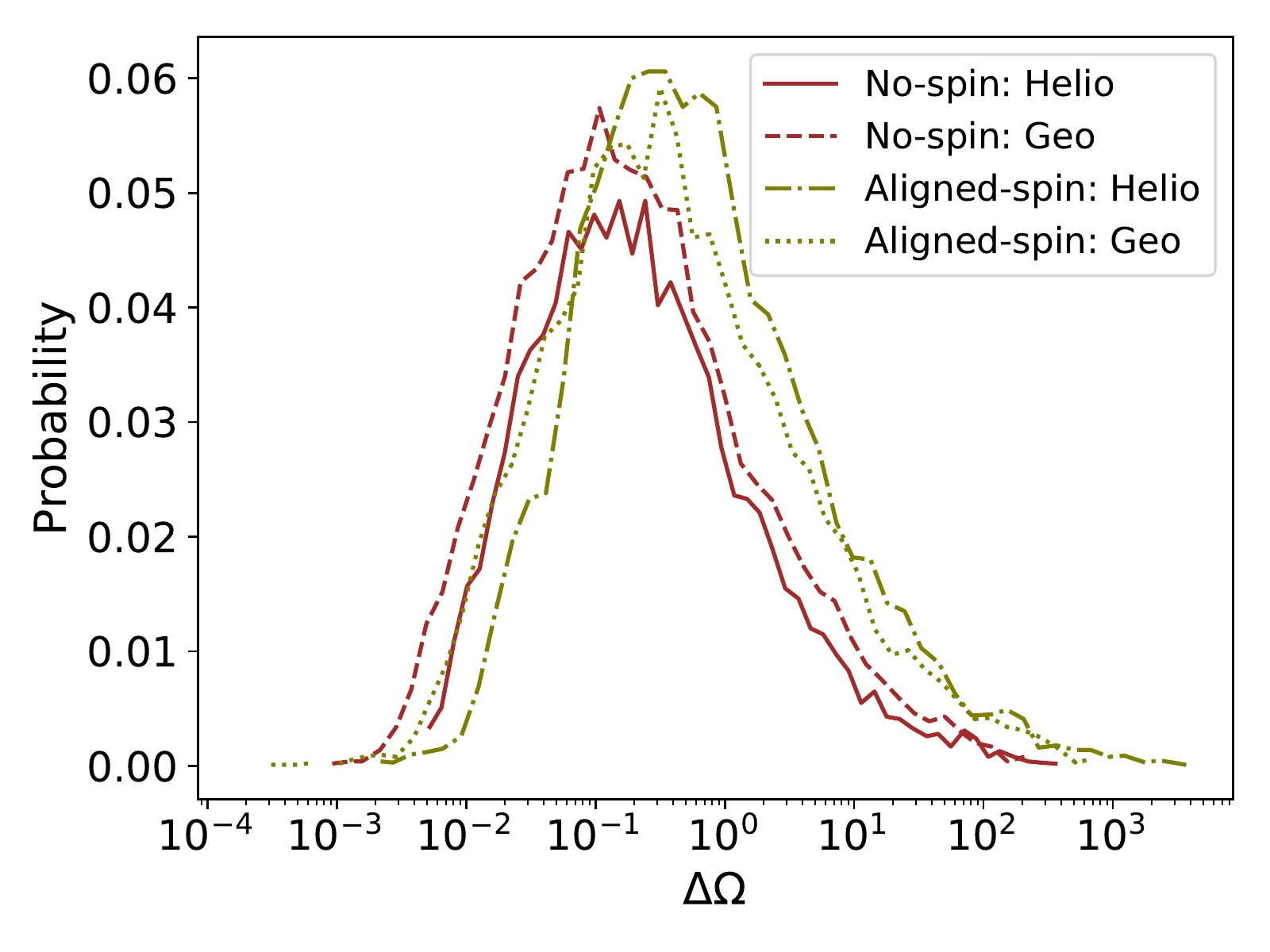}
		\label{figD_omega}
	}
	\subfloat[Part3][]
	{
		\includegraphics[width=3.1in,height=2.75in]{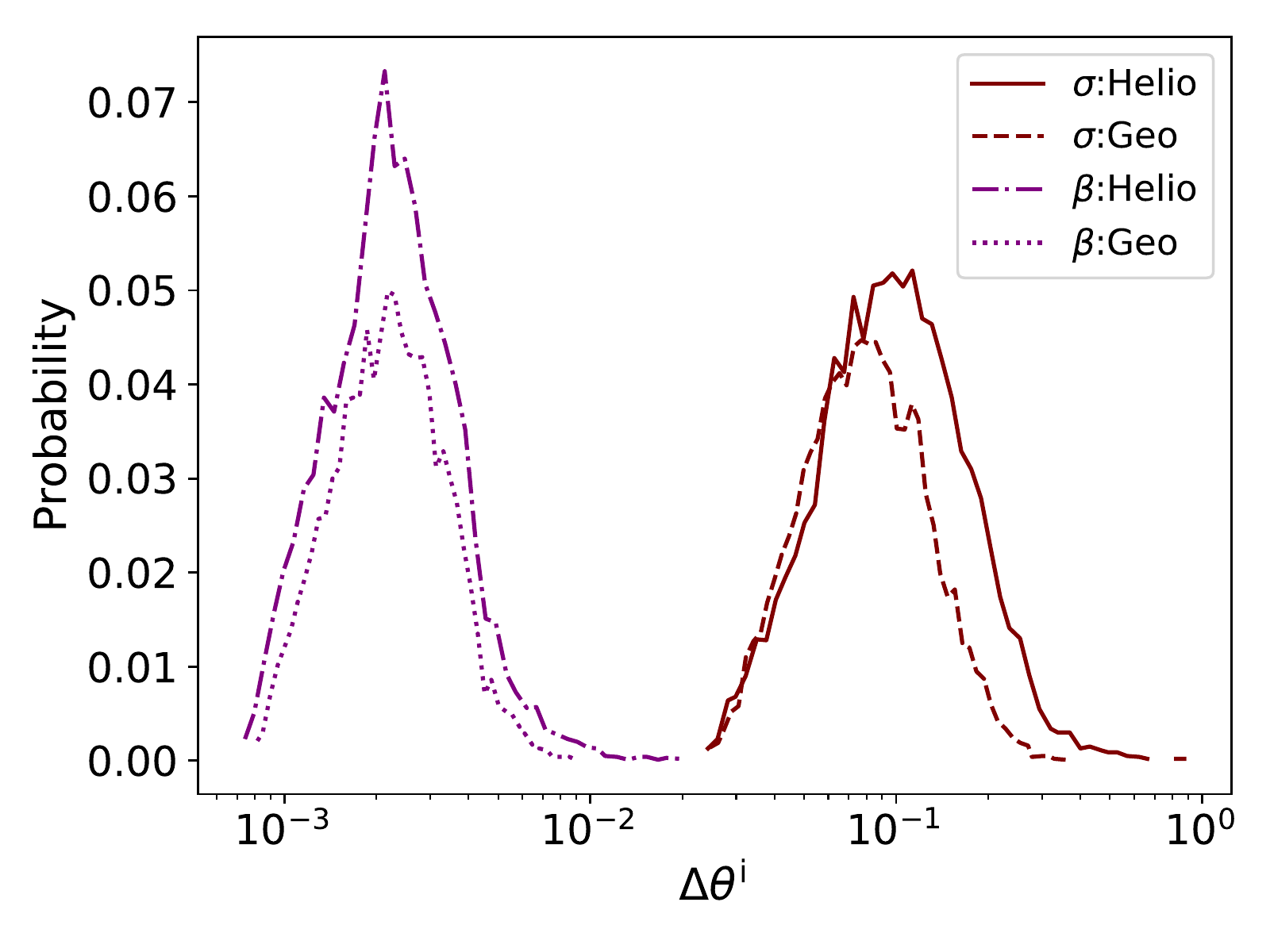}
	}
	\caption{DECIGO: Similar to Fig. \ref{figD_mc_nu}, left and right panel show the error estimates on the sky localisation (a) and the leading order spin correction terms (b), respectively.}
	\label{figD_omega_s_b}
\end{figure}

\begin{table}[ht!]
	\caption{Errors estimates for the binary coalescence parameters from DECIGO-only and joint DECIGO-ET measurements  are reported for the non-spinning case. These are calculated for BH-BH binaries with masses $30~M_{\odot}$+$40~M_{\odot}$ with the distance fixed to 3 Gpc. The fiducial values of the parameters are chosen as: $t_c=\phi_c=0$ and, choices for the angles $\bar{\theta}_{\mathrm{L}}, \bar{\theta}_{\mathrm{S}}, \bar{\phi}_{\rm{L}}, {\rm and} ~\bar{\phi}_{\mathrm{S}}$ are explained in \S \ref{nonS}. The frequency cut-offs used for calculating the Fisher matrices are mentioned in \S \ref{sec:freq_cut}. Corresponding to these cut-offs, the signal duration in the space and ground detector is $\sim$1 year and $\sim$4 seconds respectively. The first two rows of the table correspond to the helio-centric orbit (H) while the last two rows correspond to a geo-centric orbit (G) for DECIGO.}
	\vspace{0.2cm}
	\hspace{-0.35cm}
	%	\centering
	\begin{tabular}{c c c c c c c}
		\hline 
		\hline
		{\footnotesize  Detector} & 
		{\footnotesize $\Delta t_c$} & {\footnotesize $\Delta \phi_c$ } &  
		{\footnotesize  $\Delta {\cal M}/{\cal M} (\%) $ } &  
		{\footnotesize  $\Delta \nu/\nu (\%) $} & {\footnotesize  $\Delta \Omega~(\rm arcmin^2)$} &
		{\footnotesize SNR}
		\\
		\hline
		\hline 
		DECIGO (H) \quad & $1.4 \times 10^{-1}$  \quad & $1.7 \times 10^{-2}$ 
		\quad& $2.8 \times 10^{-6}$ \quad& $4.4 \times 10^{-3}$ 
		\quad& $1.8 \times 10^{-1}$ \quad& $\sim 650$ \quad\\
		\hline
		Joint (H) \quad & $2.4 \times 10^{-3}$  \quad & $1.4 \times 10^{-2}$ 
		\quad& $8.1 \times 10^{-7}$ \quad& $3.3 \times 10^{-3}$ 
		\quad& $1.8 \times 10^{-3}$ \quad&   \quad\\
		\hline
		DECIGO (G)  \quad & $1.0 \times 10^{-1}$  \quad & $1.5 \times 10^{-2}$ 
		\quad& $2.1 \times 10^{-6}$ \quad& $4.3 \times 10^{-3}$ 
		\quad& $1.6\times 10^{-1}$ \quad& $\sim 677$ \quad\\
		\hline
		Joint (G) \quad & $1.0 \times 10^{-1}$  \quad & $1.1 \times 10^{-2}$ 
		\quad& $2.0 \times 10^{-6}$ \quad& $3.2 \times 10^{-3}$ 
		\quad& $1.4 \times 10^{-1}$ \quad&   \quad\\
		\hline
	\end{tabular}
	\label{tab1}
\end{table}
\begin{table}[ht]
	\caption{Similar to Table \ref{tab1} but with scaled-DECIGO (B-DECIGO) errors shown along with joint error estimates ($S_n(f)^{\rm B-DECIGO}=10^3 S_n(f)^{\rm DECIGO}$).}
	\vspace{0.2cm}
	\hspace{-0.5cm}
	%	\centering
	\begin{tabular}{c c c c c c c}
		\hline 
		\hline
		{\footnotesize  Detector} & 
		{\footnotesize $\Delta t_c$} & {\footnotesize $\Delta \phi_c$ } &  
		{\footnotesize  $\Delta {\cal M} /{\cal M} (\%) $ } &  
		{\footnotesize  $\Delta \nu /\nu (\%)$} & {\footnotesize  $\Delta \Omega~(\rm arcmin^2)$} &
		{\footnotesize SNR}
		\\
		\hline
		\hline
		B-DECIGO (H) \quad & $4.4$  \quad & $5.5 \times 10^{-1} $ 
		\quad& $8.8 \times 10^{-5}$ \quad& $1.4 \times 10^{-1}$ 
		\quad& $1.8 \times 10^{2}$ \quad& $\sim 20$ \quad\\
		\hline
		Joint (H) \quad & $6.3 \times 10^{-2}$  \quad & $1.2 \times 10^{-1}$ 
		\quad& $2.5 \times 10^{-5}$ \quad& $4.1 \times 10^{-2}$ 
		\quad& $1.3 $ \quad&   \quad\\
		\hline
		\hline
		B-DECIGO (G) \quad & $3.3$  \quad & $4.9 \times 10^{-1} $ 
		\quad& $6.7 \times 10^{-5}$ \quad& $1.3 \times 10^{-1}$ 
		\quad& $1.6 \times 10^{2}$ \quad& $\sim 22$ \quad\\
		\hline
		Joint (G) \quad & $2.7$  \quad & $1.0 \times 10^{-1}$ 
		\quad& $5.5 \times 10^{-5}$ \quad& $4.1 \times 10^{-2}$ 
		\quad& $1.1 \times 10^{2}$ \quad&   \quad\\
		\hline
	\end{tabular}
	\label{tab2}
\end{table}
Now we tabulate and visualize the expected errors in the parameter estimation 
when we combine measurements of ground and space based detectors, ET and DECIGO. 
As mentioned earlier, we have considered a BH-BH binary systems with component masses
$30~M_{\odot}+40~M_{\odot}$ and with the distance fixed at 3 Gigaparsecs (Gpc).
To obtain the error estimates we uniformly distribute $10^4$ BH-BH sources over the sky. $\bar{\phi}_{\mathrm{S}}$ and $\bar{\phi}_{\mathrm{L}}$ are randomly generated in the range $\left[0, 2\pi \right]$ and $\cos \bar{\theta}_{\mathrm{S}}$ and 
$\cos \bar{\theta}_{\mathrm{L}}$ are randomly generated in the range $\left[-1, 1 \right]$. After computing the parameter errors for each such system, we group them into bins in a logarithmic distribution following Berti et.al., \cite{berti}. A source is assumed to belong to the $j$th bin if the error on some parameter $\theta$ satisfies
\begin{equation*}
\left[ \ln(\theta_{\min})+\frac{(j-1)[\ln(\theta_{\max})-\ln(\theta_{\min})]}{N_{\mathrm{bins}}} \right] < \ln(\theta) 
                               <\left[ \ln(\theta_{\min})+\frac{j[\ln(\theta_{\max})-\ln(\theta_{\min})]}{N_{\mathrm{bins}}} \right],
\end{equation*}
where $j=1,2,3...N$. Here $N$ is the total number of bins which we fix to 50. Once the errors are binned using the above relation, binaries in each bin are normalized (by dividing sources in each bin with the total number of binaries) to get a probability distribution of the error. Plots obtained from these histograms for each parameter error are shown in Fig. \ref{figD_mc_nu} and \ref{figD_omega}.

%%%%%%%%%%%%%
\begin{figure}[ht]
	\hspace{-0.7cm}
		\subfloat[Part 3][]
		{
			\includegraphics[width=3.1in,height=2.7in]{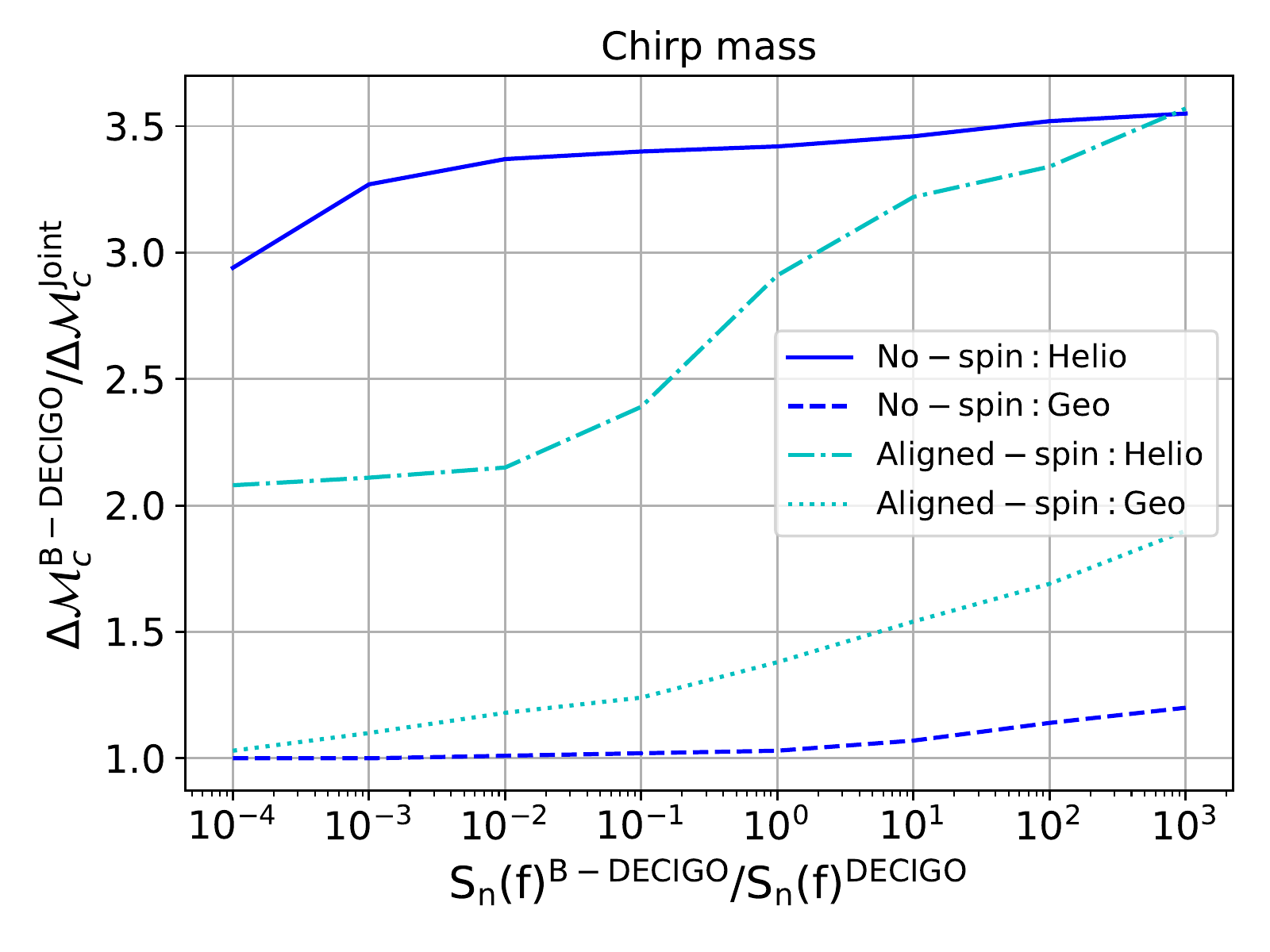}
		}
		\subfloat[Part3][]
		{
			\includegraphics[width=3.1in,height=2.7in]{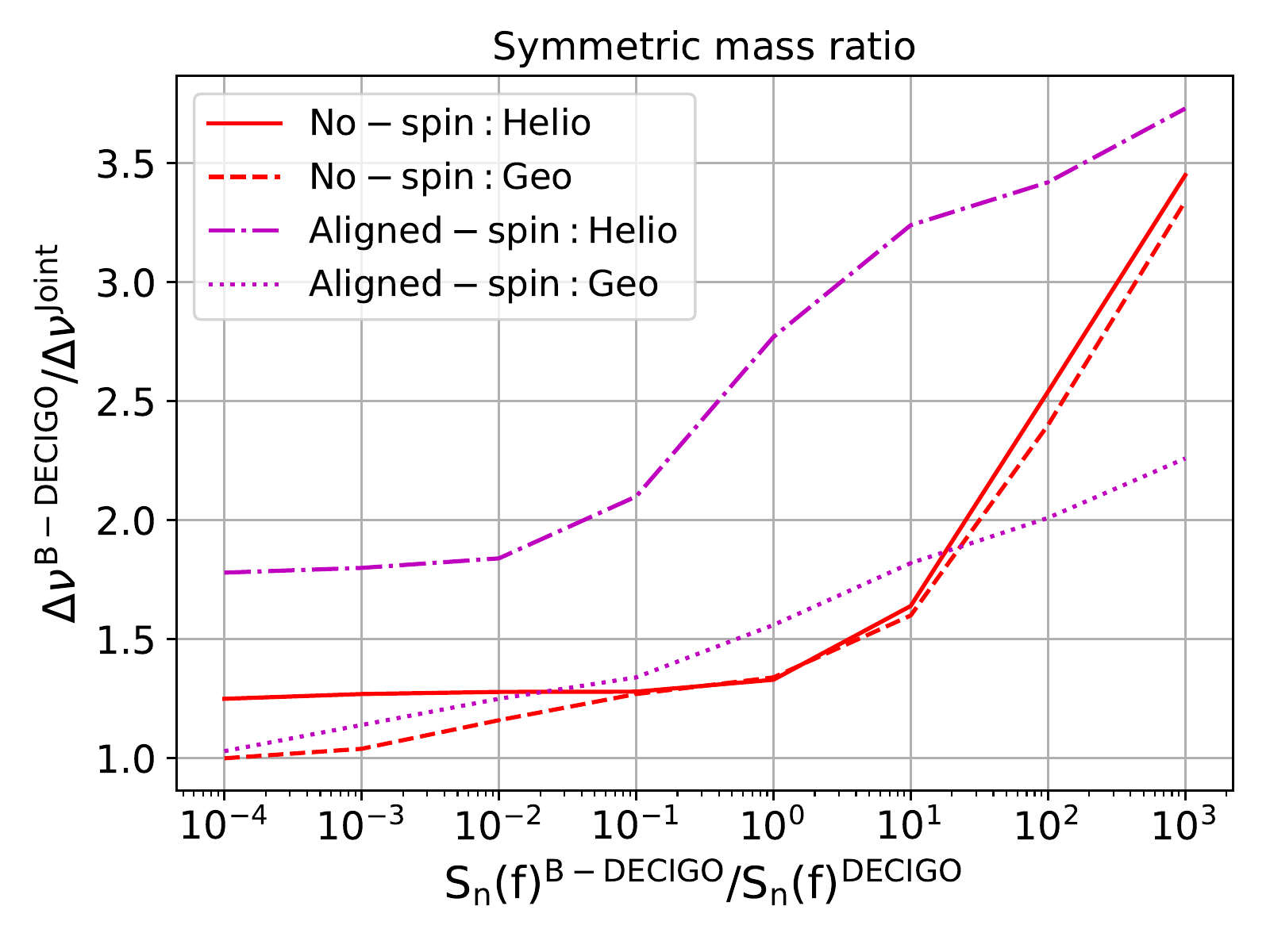}
		}
		\caption{The left panel shows the variation in the error estimates of the chirp mass for the no-spin (blue solid/dashed curves) and spin-aligned (cyan dot-dashed/dotted curve) cases, obtained by varying the noise sensitivity of DECIGO like mission (B-DECIGO) in helio-centric and geo-centric orbit configurations. The right panel shows similar results obtained for the symmetric mass ratio. The sensitivity is varied according to Eq. \eqref{sdecigo} and the ratio of error estimates for the B-DECIGO-only to B-DECIGO-ET joint measurements are plotted on the y-axis.
		}
		\label{fig_var_mc_nu}

\end{figure}
\begin{figure}[h]
	\hspace{-0.7cm}
		\subfloat[Part 3][]
		{
			\includegraphics[width=3.1in,height=2.7in]{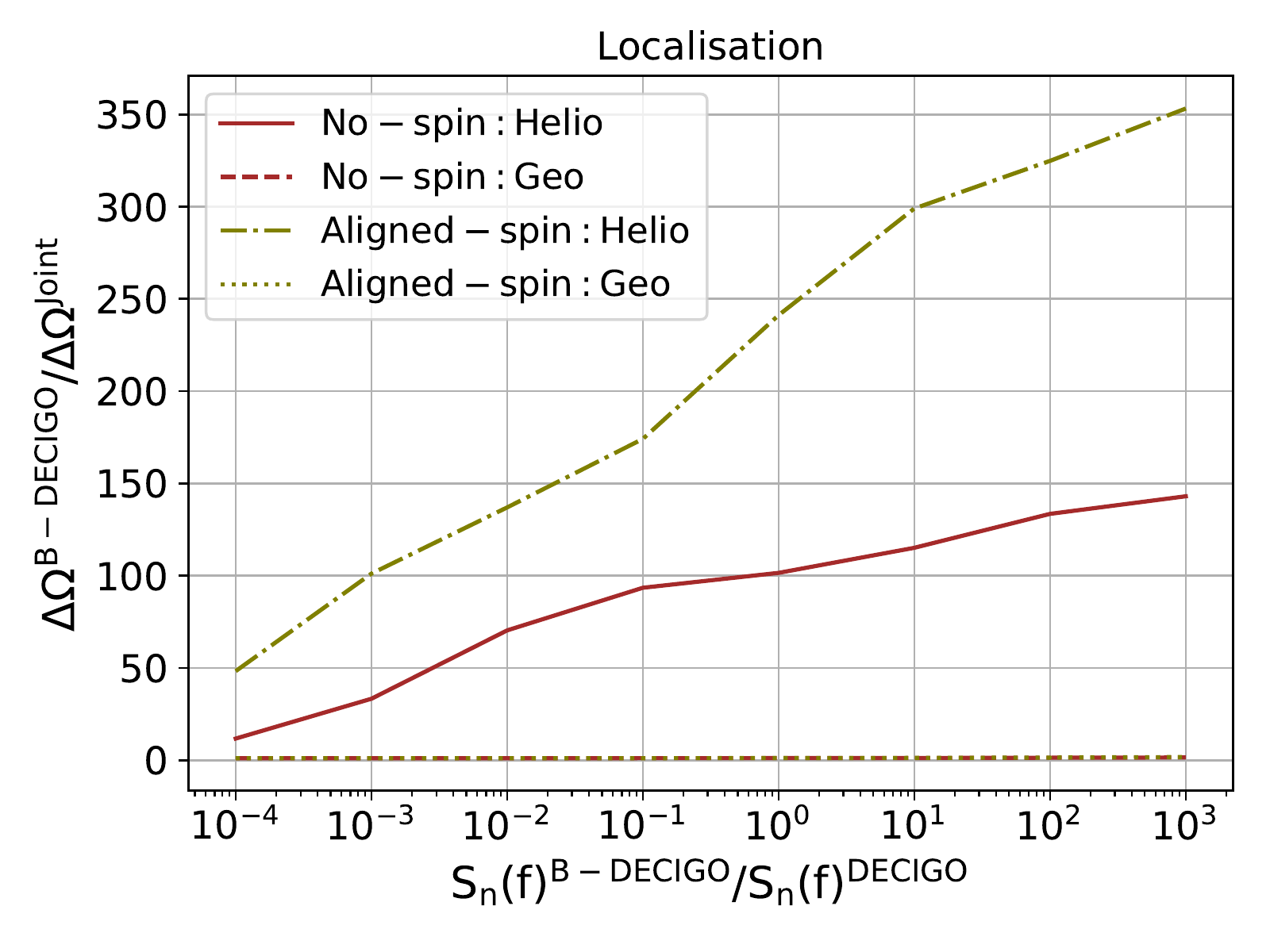}
			\label{fig_var_omega}
		}
		\subfloat[Part3][]
		{
			\includegraphics[width=3.1in,height=2.7in]{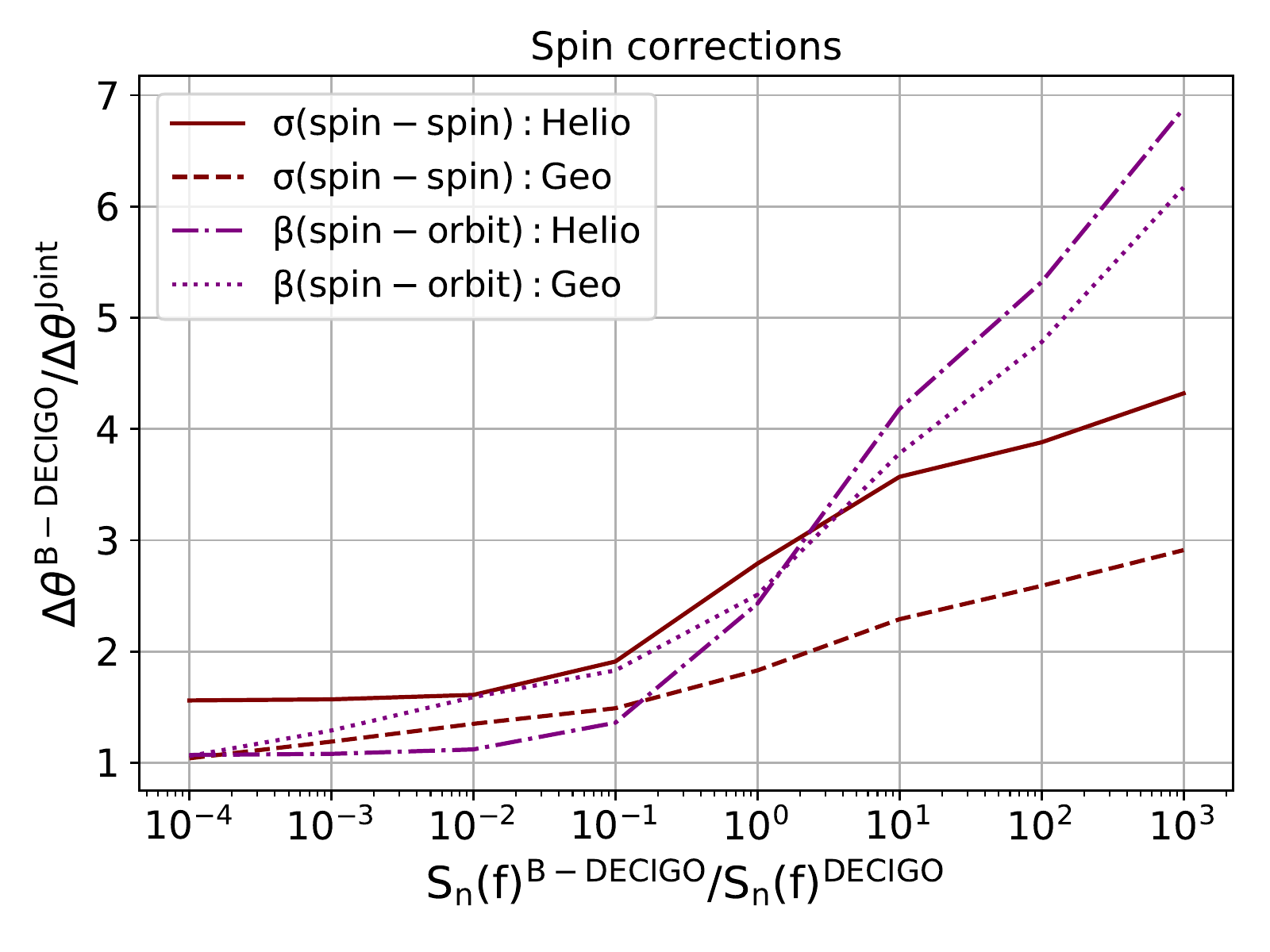}
		}
		\caption{The left panel shows the variation in the error estimates of the sky localisation for the no-spin (brown solid/dashed curves) and spin-aligned (olive dot-dashed/dotted curve) cases, obtained by varying the noise sensitivity of DECIGO like mission (B-DECIGO) in helio-centric and geo-centric orbit configurations. The right panel shows similar results obtained for the leading order spin correction terms $\sigma$ and $\beta$. The sensitivity is varied according to Eq. \eqref{sdecigo} and the ratio of the error estimates for the B-DECIGO-only to B-DECIGO-ET joint measurements is plotted on the y-axis.
		}
		\label{fig_var_omega_s_b}

\end{figure}
We report the results of combining the ET-DECIGO measurements in Table \ref{tab1} and we see that although the improvement for most parameters is not significant (DECIGO dominates the error budget) the localisation is improved from $\sim$ 0.18 arcmin$^2$ to $\sim$ 1.8 $\times 10^{-3}$ arcmin$^2$ (first two rows). Similar results are obtained for the case of geo-centric DECIGO orbit (bottom two rows), but unlike the helio-centric case, here we see almost no improvement in localisation.  Results obtained for B-DECIGO are shown in Figs. \ref{fig_var_mc_nu} and \ref{fig_var_omega} where the variation in the error estimates are plotted with varying B-DECIGO sensitivity. We cut off this curve where the SNR for B-DECIGO falls below $\sim$10 (this happens when $S_n(f)^{\rm B-DECIGO}$ / $S_n(f)^{\rm DECIGO} \sim 10^{3}$). In Table \ref{tab2} we report the error estimates for the case where we have maximum synergy between ET and B-DECIGO (with SNR threshold at $\sim$10). We see that one can obtain better estimates for all the parameters considered, and the maximum improvement is seen for the time of coalescence and localisation in the helio-centric DECIGO case. The localisation from B-DECIGO of $\sim$ $1.8 \times 10^2$ arcmin$^2$ is reduced to $\sim 1.3$ arcmin$^2$ when the ET measurements are included (top two rows in Table. \ref{tab2}). 
\subsection{Spin-(anti)aligned BH-BH binary}
In this section we comment on the results for the spinning (non-precessing) case. As mentioned earlier, we have a 10 dimensional parameter space in this case. We forecast the parameter errors in a similar way as in the non-spinning case by constructing the Fisher matrix and subsequently inverting it to obtain the parameter errors. To include all the spin corrections (instead of just the leading order $\beta$ and $\sigma$) one can write the correction terms using the variables $s_1 = \vec \chi_s \cdot  \hat{{\rm L}}$ and 
$s_2 = \vec \chi_a \cdot  \hat{{\rm L}}$. The spin corrections are then given as:
\begin{eqnarray}
\beta &=& \left(\frac{113}{12} -\frac{19}{3} \nu \right) s_ 1+ \frac{113}{12} s_2 \delta  \ , \nonumber \\
\sigma &=&  \frac{81}{16}\left(s_1^2 + s_2^2 \right) - \frac{1}{4} \nu s_1^2 -20 \nu s_2^2 +\frac{81}{8} s_1 s_2 \delta  \ , \nonumber \\
\gamma &=&  \left ( \frac{732985}{2268} - \frac{24260}{81} \nu - \frac{340}{9} \nu^2 \right) s_1 + \left(\frac{732985}{2268} + \frac{140}{9} \nu \right) s_2 \delta \ , \nonumber \\
\xi &=&  \left( \frac{75}{2} -\frac{74}{3}\nu \right) \pi s_1 + \frac{75}{2} \pi s_2 \delta \ , \nonumber \\
\zeta &=&  \left (\frac{130325}{756} - \frac{1575529}{2592} \nu + \frac{341753}{864} \nu^2 -\frac{10819}{216} \nu^3 \right) s_1  +  \left(\frac{130325}{756} - \frac{796069}{2016} \nu + \frac{100019}{864} \nu^2 \right) s_2 \delta , \nonumber \\
\hfill
\end{eqnarray}
where $\delta = (m_1-m_2)/(m_1+m_2) = \sqrt{1-4 \nu}$.  
%We find that if we account for all the spin corrections, the Fisher matrix is very ill behaved for many of the inclination angles and orientations. Hence it is not straight-forward to obtain the parameter errors. Degeneracy between parameters could be one reason for such non-invertible matrices. 
In this work we only report the results for the leading order spin-spin ($\sigma$) and spin-orbit ($\beta$) corrections. 
%We will address the higher order spin corrections and precession in a future work.
%%%%%%%------------------------------------

We find that the estimates for the chirp mass, symmetric mass ratio and the sky localisation are worsened, compared to the non-spinning case, if we account for $\sigma$ and $\beta$ (dot-dashed and dotted curves in Figs. \ref{figD_mc_nu} and \ref{figD_omega_s_b}). We further note that although the cutoff frequency depends also on spins, here it is determined by the total mass only and our results may change if the effect of spins on $f_{\rm LSO}$ is taken into account.
\begin{table}[ht!]
	\caption{Errors estimates for the binary coalescence parameters from DECIGO-only and joint DECIGO-ET measurements  are reported for the aligned-spin case. These are calculated for BH-BH binaries with masses $30~M_{\odot}$+$40~M_{\odot}$ with the distance fixed to 3 Gpc. The fiducial values of the parameters are chosen as: $t_c=\phi_c=\sigma=\beta=0$ and, choices for the angles $\bar{\theta}_{\mathrm{L}}, \bar{\theta}_{\mathrm{S}}, \bar{\phi}_{\rm{L}}, {\rm and} ~\bar{\phi}_{\mathrm{S}}$ are explained in \S \ref{nonS}. The frequency cut-offs used for calculating the Fisher matrices are mentioned in \S \ref{sec:freq_cut}. Corresponding to these cut-offs, the signal duration in the space and ground detector is $\sim$1 year and $\sim$4 seconds respectively. The first two rows correspond to the helio-centric orbit (H) while the last two rows correspond to a geo-centric orbit (G) for DECIGO.}
	\vspace{0.2cm}
	\hspace{-1.7cm}
	%\resizebox{\textwidth}{!}{%
	%\centering
	\footnotesize
	\begin{tabular}{c c c c c c c c c}
		\hline 
		\hline
		{\footnotesize  Detector} & 
		{\footnotesize $\Delta t_c$} & {\footnotesize $\Delta \phi_c$ } &  
		{\footnotesize  $\Delta {\cal M}/{\cal M}(\%) $ } &  
		{\footnotesize  $\Delta \nu/\nu (\%) $} & {\footnotesize  $\Delta \Omega~(\rm arcmin^2)$}
		& {\footnotesize  $\Delta \sigma$} & {\footnotesize  $\Delta \beta$} &
		{\footnotesize SNR}
		\\
		\hline
		\hline
		
		DECIGO (H) \quad & $4.2 \times 10^{-1}$  \quad & $1.6 \times 10^{-1}$ 
		\quad& $6.2 \times 10^{-5}$ \quad& $2.4 \times 10^{-1}$ 
		\quad& $4.8 \times 10^{-1}$ \quad& $9.6 \times 10^{-2}$ \quad& $2.2 \times 10^{-3}$
		\quad& $\sim 650$ \quad\\
		Joint (H)\quad & $2.6 \times 10^{-3}$  \quad & $5.2 \times 10^{-2}$ 
		\quad& $2.1 \times 10^{-5}$ \quad& $8.6\times 10^{-2}$ 
		\quad& $2 \times 10^{-3}$ \quad& $3.4 \times 10^{-2}$ \quad& $9.1 \times 10^{-4}$
		\quad&   \quad\\
		\hline
		DECIGO (G)\quad & $2.2 \times 10^{-1}$  \quad & $1.4 \times 10^{-1}$ 
		\quad& $4.4 \times 10^{-5}$ \quad& $1.8 \times 10^{-1}$ 
		\quad& $3.1   \times 10^{-1}$ \quad& $7.8 \times 10^{-2}$ \quad& $2.2 \times 10^{-3}$
		\quad& $\sim 678$ \quad\\
		Joint (G) \quad & $2.1 \times 10^{-1}$  \quad & $4.9 \times 10^{-2}$ 
		\quad& $3.2 \times 10^{-5}$ \quad& $1.1 \times 10^{-1}$ 
		\quad& $2.8 \times 10^{-1}$ \quad& $4.2 \times 10^{-2}$ \quad& $8.7 \times 10^{-4}$
		\quad&   \quad\\
		\hline
	\end{tabular}
	\label{tab3}
\end{table}
%%%%%%%%%%%%%
\begin{table}[ht]
	\caption{Similar to Table \ref{tab3} but with scaled-DECIGO errors shown along with joint error estimates ($S_n(f)^{\rm B-DECIGO}=10^3 S_n(f)^{\rm DECIGO}$).}
	\vspace{0.2cm}
	\hspace{-1.7cm}
	%\resizebox{\textwidth}{!}{%
	%\centering
	\footnotesize
	\begin{tabular}{c c c c c c c c c}
		\hline 
		\hline
		{\footnotesize  Detector} & 
		{\footnotesize $\Delta t_c$} & {\footnotesize $\Delta \phi_c$ } &  
		{\footnotesize  $\Delta {\cal M}/{\cal M}(\%)$ } &  
		{\footnotesize  $\Delta \nu/\nu(\%)$} & {\footnotesize  $\Delta \Omega~(\rm arcmin^2)$}
		& {\footnotesize  $\Delta \sigma$} & {\footnotesize  $\Delta \beta$} &
		{\footnotesize SNR}
		\\
		\hline
		\hline
		B-DECIGO (H) \quad & $1.3 \times 10^{1}$  \quad & $4.9$ 
		\quad& $1.9 \times 10^{-3}$ \quad& $7.5 $ 
		\quad& $4.8 \times 10^{2}$ \quad& $3.0$ \quad& $7.0 \times 10^{-2}$
		\quad& $\sim 20$ \quad\\
		Joint (H) \quad & $6.5 \times 10^{-2}$  \quad & $4.3 \times 10^{-1}$ 
		\quad& $5.5 \times 10^{-4}$ \quad& $2.0$ 
		\quad& $1.4 $ \quad& $7.0 \times 10^{-1}$ \quad& $1.0 \times 10^{-2}$
		\quad&   \quad\\
		\hline
		B-DECIGO (G) \quad & $7.1 $  \quad & $4.5$ 
		\quad& $1.4 \times 10^{-3}$ \quad& $5.7$ 
		\quad& $3.3 \times 10^{2}$ \quad& $2.4$ \quad& $6.9 \times 10^{-2}$
		\quad& $\sim 22$ \quad\\
		Joint (G) \quad & $4.6$  \quad & $4.5 \times 10^{-1}$ 
		\quad& $7.3 \times 10^{-4}$ \quad& $2.5$ 
		\quad& $1.9  \times 10^{2}$ \quad& $8.5 \times 10^{-1}$ \quad& $1.1 \times 10^{-2}$
		\quad&  \quad\\
		\hline
		
	\end{tabular}
	\label{tab4}
\end{table}

The histograms for all the parameter errors are obtained in a similar manner as the non-spinning case and are shown in Figs. \ref{figD_mc_nu} and  \ref{figD_omega_s_b} (dot-dashed and dotted curves). Here also, we vary the B-DECIGO sensitivity to study the synergy between ground and space detector, and we find that similar to the non-spinning case, maximum synergy is obtained when $S_n(f)^{{\rm B-DECIGO}} = 10^3 S_n(f)^{{\rm DECIGO}}$ (with SNR threshold $>$10). For the helio-centric DECIGO case, the error estimates improve for all parameters when the measurements are combined, and again, as in the case of non-spinning binaries, we see that the maximum improvement is in time of coalescence and localisation. For ET-DECIGO pair we see that the localisation improves from $\sim 4.8 \times 10^{-1}$  arcmin$^2$ to $\sim 2 \times 10^{-3}$ arcmin$^2$ and in ET- (B-DECIGO) case it is improved from $\sim 4.8 \times 10^2$  arcmin$^2$ to $\sim$ 1.4 arcmin$^2$. For the geo-centric case we see that when the ET measurements are  combined, better constraints are obtained for all the parameters except for localisation. 

As also mentioned earlier, compared to the geo-centric DECIGO orbit, the synergy between the space and ground detector is mildly larger in the case of helio-centric DECIGO orbit. For the symmetric mass ratio and chirp mass the difference is not very remarkable (Fig. \ref{fig_var_mc_nu}). But, for the case of localisation error, this difference is quite significant as seen in Fig. \ref{fig_var_omega_s_b}(a), where $\Delta \Omega^{\rm B-DECIGO}/\Delta \Omega^{\rm Joint}\sim1$ from geo-centric DECIGO. Again, this is because compared to the geo-centric DECIGO, the distance between the ground and space detector is larger in the case of the helio-centric DECIGO and this helps to improve localisation.
%%%%%%%%%%%%%%%%%%%%%%%%%%%%%%%%%%%%%%%
\section{Implications}\label{sec:impli}
In this paper we have assessed the expected synergy effects between ground and space based detectors, in the determination of binary coalescence parameters. For this, we study the estimated errors on parameters of these systems by considering $30$ $M_{\odot}$ + $40$ $M_{\odot}$ BH-BH binaries. This mass range corresponds to the first GW detection (GW150914), and is compatible with the mass ranges of subsequent BH-BH detections. 

We studied two cases: non spinning BH-BH binaries and spin-aligned (non-precessing) BH-BH binaries, with two different detector configuration for DECIGO (helio-centric and geo-centric). For the helio-centric DECIGO orbit we found that combining measurements with ET gave us better error estimates for all parameters and the gain was most significant for the time of coalescence and the localisation of the source. This improvement in localisation is very crucial for the future of GW astronomy, as it gives us a chance to identify the host galaxy. We did not find large synergy between the space and ground based detectors for the geo-centric DECIGO orbit case.

Binary BH mergers are not expected to be accompanied by an electro-magnetic event but if the localisation of the source is good enough, one can expect to identify the host galaxies through galaxy catalogs or through dedicated survey of the localisation area. The large synoptic survey telescope which will begin science operation around 2022, will survey nearly 18,000 square degrees of the sky \cite{lsst} and the proposed BigBOSS survey is an all sky galaxy redshift survey, spanning redshift from $0.2<z<3.5$ \cite{bigb}. Hence, there is hope that by the time we achieve $\sim$sub-arcmin$^2$ accuracy on localisation, as is seen for the ET+(B-)DECIGO joint measurements, such surveys would have determined redshifts for a large fraction of the possible host galaxies (in a good fraction of the sky). Note that there are selection effects in all electro-magnetic surveys as there are mass (luminosity) cuts, and we are making an optimistic assumption that host galaxies would be seen by these surveys. In case the merger happens in galaxies that are not observed by these surveys, we would not be able to identify the host galaxy even if the sky localisation is small. 

Identifying the host galaxies of GW events would further give us a chance to do cosmological studies with GW observations. Here we highlight how this may be achieved by doing joint measurements. If we make some simple estimates for milky-way type galaxies, similar to those quoted in \cite{nissanke}, we find that combining measurements from DECIGO and ET for the aligned-spin case reduces the number of possible host galaxies in the localisation region from $\sim$9 to $ \sim 1$, and for B-DECIGO+ET this number reduces from $\sim 8.6 \times 10^3$ galaxies to $\sim$24 galaxies (both numbers for helio-centric case with $S_n(f)^{\rm B-DECIGO}=10^3 S_n(f)^{\rm DECIGO}$). To get these numbers, we have assumed the Schechter luminosity function \cite{schechter} which provides a description of the density of galaxies as a function of their luminosity: $\rho_{gal}(x)dx = \phi^{*}x^a\exp^{-x}dx$, where $x=L/L^*$ and $L^*$ is some characteristic luminosity where the power law character of the function truncates. After making the same assumptions for $\phi^*$, $a$ and $L^*$ (from B-band measurements) as in \cite{nissanke}, we get the galaxy density above $x_{1/2}$ (for $a = -1.07$, half of the luminosity density is contributed by galaxies with $x_{1/2}> 0.626$) as $2.35 \times 10^{-3}$ Mpc$^{-3}$. To get the numbers quoted above we multiply this density by a volume element $\Delta V = (4/3) ~\pi ~D_L^3 (\Delta \Omega/4 \pi)$. These numbers may further decrease if we take into account the distance estimate and the corresponding error since the volume element would significantly decrease ($\Delta V = \Delta \Omega~ D_L^3 ~(2 \times \Delta D_L/D_L)$) when we have good distance estimates. As an example we find that for the spin-aligned case, $\Delta D_L/D_L \sim 6.4 \%$ for  (helio-centric) B-DECIGO measurements of binaries located at 3 Gpc. Incorporating this distance error estimate we find that the number of galaxies in the volume element reduces from $\sim 3000$ for B-DECIGO measurements to $\sim 9$ galaxies for B-DECIGO+ET measurements. For BH-BH binaries at a distance $\sim$400 Mpc (GW150914) these numbers would further reduce by a factor of $\sim$1000. Namely, at lower distances we can expect the number of potential host galaxies to reduce to just a few, even in the case of B-DECIGO+ET measurements.

In cases where there are many galaxies within the sky localisation region, one can use the distance information obtained from GW measurements with a well motivated distance-redshift relation to rule out those galaxies which are at the right position on the sky but have significantly different redshifts \cite{cutler_holz}. We stress again that though these numbers may not be very robust, they are indicative of what can be observed in the future and this is very encouraging for the future of GW astronomy. 

In this work we only accounted for the leading order spin-spin and spin-orbit corrections to the phase of the GW waveform. Also, for simplicity we considered spin-aligned (non-precessing) waveforms. Including precession is very important for unequal-mass systems (NS-BH binaries) as unequal-mass systems precess more than equal-mass systems. There are many studies in the literature that explore the effect of including precession on the parameter estimation (for space-based detectors) \cite{Lang06,Vecc04,Lang11}, and they have found that including precession can improves parameter estimation by breaking parameter degeneracies. Including eccentricity may also effect the parameter estimation (\cite{yagi_tanaka}), but since we focus on very late phase of the inspiral in the space based band, we have neglected the effect of eccentricity in our study, and used waveforms for quasi-circular orbits. In a recent paper, authors studied multi-band measurements of non-spinning binary NS and aligned-spin BH-BH systems with B-DECIGO/LISA and advanced LIGO/ET detectors. Neglecting the sky localisation information, they focused on the parameter estimation accuracy of  mass, NS Love numbers, and the BH spins \cite{nakano}. One can also study the improvement in localisation of GW sources when a ground based detector network is considered instead of a single detector. It will be interesting to study the modifications in our results due to these effects (higher spin corrections, precession, eccentricity, ground based detector network) and we will consider them in future publications.
\section{Acknowledgment}
R. N is an international research fellow of the Japan Society for the Promotion of Science (JSPS) and acknowledges support from JSPS grant No. 16F16025. T. T acknowledges support in part by MEXT Grant-in-Aid for Scientific Research on Innovative Areas, Nos. 17H06357 and 17H06358, and by Grant-in-Aid for Scientific Research Nos. 26287044 and 15H02087. Authors thank Chandra Kant Mishra and Nathan Johnson-McDaniel for discussions.
%\end{acknowledgements}
\appendix

\section{Detector response to inspiraling binary signals}\label{app_detResponse}
In this section we briefly layout the expressions used to obtain the GW waveform. The beam pattern function in Eq. \eqref{Apol} are given by:
\begin{eqnarray}
F_{\mathrm{I}}^{+}(\theta_{\mathrm{S}},\phi_{\mathrm{S}},\psi_{\mathrm{S}}) 
                &=&\frac{1}{2}(1+\cos^2 \theta_{\mathrm{S}}) \cos(2\phi_{\mathrm{S}}) \cos (2\psi_{\mathrm{S}})
                  -\cos(\theta_{\mathrm{S}}) \sin(2\phi_{\mathrm{S}}) \sin(2\psi_{\mathrm{S}}),  \\
F_{\mathrm{I}}^{\times}(\theta_{\mathrm{S}},\phi_{\mathrm{S}},\psi_{\mathrm{S}})
                &=&\frac{1}{2}(1+\cos^2 \theta_{\mathrm{S}}) \cos(2\phi_{\mathrm{S}}) \sin (2\psi_{\mathrm{S}})
                  +\cos(\theta_{\mathrm{S}}) \sin(2\phi_{\mathrm{S}}) \cos(2\psi_{\mathrm{S}}). \label{beam-pattern}
\end{eqnarray} 
Here $(\theta_{\mathrm{S}},\phi_{\mathrm{S}})$ represents the direction of the source in the detector frame and $\psi_{\mathrm{S}}$ is the polarisation angle defined as 
\begin{equation}
\tan\psi_{\mathrm{S}}    =\frac{\hat{\bm{L}}\cdot\hat{\bm{z}}-(\hat{\bm{L}}\cdot\hat{\bm{N}})(\hat{\bm{z}}\cdot\hat{\bm{N}})}{\hat{\bm{N}}\cdot(\hat{\bm{L}}\times\hat{\bm{z}})}. 
%\label{tanpsi}
\end{equation}
DECIGO, since it has three arms, corresponds to having two individual detectors. Therefore, it is possible to measure both polarisations with one detector. One can reduce DECIGO to two independent interferometers with an equilateral triangle shape. If such an equilateral triangle is placed symmetrically inside the 90$\degree$ interferometer then the beam-pattern functions for the two detectors are the same as for a single detector, (except for the factor $\sqrt{3}/2$). The beam pattern function for the second detector output is given by 

\begin{eqnarray}
F_{\mathrm{II}}^{+}(\theta_{\mathrm{S}},\phi_{\mathrm{S}},\psi_{\mathrm{S}})&=&F_{\mathrm{I}}^{+}(\theta_{\mathrm{S}},\phi_{\mathrm{S}}-\pi/4,\psi_{\mathrm{S}}), \\
F_{\mathrm{II}}^{\times}(\theta_{\mathrm{S}},\phi_{\mathrm{S}},\psi_{\mathrm{S}})&=&F_{\mathrm{I}}^{\times}(\theta_{\mathrm{S}},\phi_{\mathrm{S}}-\pi/4,\psi_{\mathrm{S}}).
\end{eqnarray}

\section{Detector orbit}\label{app_detOrbit}
While performing parameter estimation we take the direction of the source $(\bar{\theta}_{\mathrm{S}},\bar{\phi}_{\mathrm{S}})$ and the direction of the orbital angular momentum $(\bar{\theta}_{\mathrm{L}},\bar{\phi}_{\mathrm{L}})$, both in the solar barycentric frame.
Therefore we need to express the waveforms (especially $\hat{\bm{L}}\cdot\hat{\bm{N}}$ and the beam-pattern functions $F_{\alpha}^{+}$ and $F_{\alpha}^{\times}$ which appear in Eqs.~(\ref{Apol})-(\ref{phipol2})) in terms of the barred coordinates $\bar{\theta}_{\mathrm{S}},\bar{\phi}_{\mathrm{S}},\bar{\theta}_{\mathrm{L}}$ and $\bar{\phi}_{\mathrm{L}}$.
We express $\theta_{\mathrm{S}}(t)$, $\phi_{\mathrm{S}}(t)$ and other required quantities in terms of the barred coordinates, for the different detectors in the following section.

\subsubsection*{ET}
In the expressions below $\delta=39\degree$ specifies the location of the detector on the Earth (latitude), $\epsilon = 23.4 \degree$ is the inclination of Earth's equator with respect to the ecliptic plane, $R_E$ is the radius of the Earth, $R_{AU}$ is the astronomical unit, $\phi_E = 2 \pi t [ (1/T_E)-(1/T) ]$ and $\bar{\phi}(t) = 2 \pi t/T$, where $T$ and $T_E$ correspond to 1 year and 1 day respectively.
\begin{eqnarray}
\cos \theta_{\mathrm{S}}(t)&=&\cos \bar{\theta}_{\mathrm{S}} \left(\cos \delta \cos \epsilon - \sin \delta \sin \epsilon \cos \phi_E \right) \nonumber \\ &+& \sin \bar{\theta}_{\mathrm{S}}  \left[\cos \bar{\phi}_{\mathrm{S}}  \left(\cos \delta \cos \epsilon + \sin \delta \cos \epsilon \cos \phi_E \right) -\sin \bar{\phi}_{\mathrm{S}}  \cos \delta \sin \phi_E  \right]  ,  \\
\phi_{\mathrm{S}}(t)&=&\tan^{-1}
\left( \frac{y_s}{x_s} \right),
\end{eqnarray}
where 
\begin{eqnarray}
x_s&=& \cos \bar{\theta}_{\mathrm{S}} \left(\sin \delta \cos \epsilon - \cos \delta \sin \epsilon \cos \phi_E \right) \nonumber \\ &&+ \sin \bar{\theta}_{\mathrm{S}}  \left[ \cos \bar{\phi}_{\mathrm{S}} \left(\sin \delta \sin \epsilon + \cos \delta \cos \epsilon \cos \phi_E \right) 
+ \sin \bar{\phi}_{\mathrm{S}}  \cos \delta \sin \phi_E  \right], \nonumber  \\
y_s&=& \cos \bar{\theta}_{\mathrm{S}} \sin \epsilon \sin \phi_E  + \sin \bar{\theta}_{\mathrm{S}}  \left[ - \cos \bar{\phi}_{\mathrm{S}}
 \cos \epsilon \sin \phi_E  +\sin \bar{\phi}_{\mathrm{S}} \cos \phi_E  \right].
\end{eqnarray}
The polarisation angle $\psi_{\mathrm{S}}$ is given as:
\begin{equation}
\tan\psi_{\mathrm{S}}=\frac{\hat{\bm{L}}\cdot\hat{\bm{z}}-(\hat{\bm{L}}\cdot\hat{\bm{N}})(\hat{\bm{z}}\cdot\hat{\bm{N}})}
{\hat{\bm{N}}\cdot(\hat{\bm{L}}\times\hat{\bm{z}})}.
\end{equation}
where  $\hat{\bm{z}}\cdot\hat{\bm{N}}=\cos\theta_{\mathrm{S}}$ and since in this work we neglect the spin precessional effects, $\hat{\bm{L}}$ is a constant. $\hat{\bm{L}}\cdot\hat{\bm{z}}$, $\hat{\bm{L}}\cdot\hat{\bm{N}}$, and $\hat{\bm{N}}\cdot(\hat{\bm{L}}\times\hat{\bm{z}})$ are given in terms of the barred coordinates by:
\begin{eqnarray}
\hat{\bm{L}}\cdot\hat{\bm{z}}&=&\cos \bar{\theta}_{\mathrm{L}} \left(\cos \delta \cos \epsilon - \sin \delta \sin \epsilon \cos \phi_E \right) \nonumber \\ &&+ \sin \bar{\theta}_{\mathrm{L}} \left[\cos \bar{\phi}_{\mathrm{L}} \left(\cos \delta \sin \epsilon + \sin \delta \cos \epsilon \cos \phi_E \right) +\sin \bar{\phi}_{\mathrm{L}} \sin \delta \sin \phi_E \right]    \nonumber \label{lz} \\
\hat{\bm{L}}\cdot\hat{\bm{N}}&=&\cos\bar{\theta}_{\mathrm{L}}\cos\bar{\theta}_{\mathrm{S}}
+\sin\bar{\theta}_{\mathrm{L}}\sin\bar{\theta}_{\mathrm{S}}\cos(\bar{\phi}_{\mathrm{L}}-\bar{\phi}_{\mathrm{S}}), \nonumber  \label{ln} \\
\hat{\bm{N}}\cdot(\hat{\bm{L}}\times\hat{\bm{z}})&=&\sin\bar{\theta}_{\mathrm{L}}\sin\bar{\theta}_{\mathrm{S}}
\sin(\bar{\phi}_{\mathrm{L}}-\bar{\phi}_{\mathrm{S}}) \left(\cos \delta \cos \epsilon + \sin \delta \sin \epsilon \cos \phi_E \right)   \\
&&+\sin \delta \sin \phi_E (\cos\bar{\theta}_{\mathrm{S}}\cos\bar{\phi}_{\mathrm{L}}\sin \bar{\theta}_{\mathrm{L}} -\cos\bar{\theta}_{\mathrm{L}}\cos\bar{\phi}_{\mathrm{S}}\sin \bar{\theta}_{\mathrm{S}})
 \nonumber \\
 && + \left(\cos \delta \sin \epsilon + \sin \delta \cos \epsilon \cos \phi_E \right)  (\cos\bar{\theta}_{\mathrm{L}}\sin \bar{\phi}_{\mathrm{S}} \sin \bar{\theta}_{\mathrm{S}} -\sin \bar{\theta}_{\mathrm{L}} \sin \bar{\phi}_{\mathrm{L}}\cos \bar{\theta}_{\mathrm{S}}) \nonumber 
\label{nlz}
\end{eqnarray}
The Doppler phase which contains angular information is given by:
\begin{eqnarray}
\phi_D &=& 2 \pi f \left\lbrace R_{\rm AU} \sin \bar{\theta}_{\mathrm{S}} \cos[\bar{\phi}(t)-\bar{\phi_{\mathrm{S}}}] + R_E \cos \theta_{\mathrm{S}}(t) \right \rbrace
\end{eqnarray}

\subsubsection*{Helio-centric DECIGO}
For details on the the detector configuration we refer the readers to  \cite{Seto:2001qf}. In the following expressions, $R_{AU}$ is the astronomical unit and $\bar{\phi}(t) = 2 \pi t/T$ where $T$ is equal to 1 year. We assume that the detector follows the same helio-centric orbit as the Earth, but keeping its position $\pi/9$ radians behind it. 
The location of the binary source is written in terms of the barred coordinates as:
\begin{eqnarray}
\cos \theta_{\mathrm{S}}(t)&=&\frac{1}{2}\cos \bar{\theta_{\mathrm{S}}}
-\frac{\sqrt{3}}{2}\sin \bar{\theta_{\mathrm{S}}} \cos[\bar{\phi}(t)-\bar{\phi_{\mathrm{S}}}],  \\
\phi_{\mathrm{S}}(t)&=&\tan^{-1}
\left( \frac{\sqrt{3}\cos{\bar{\theta_{\mathrm{S}}}}
	+\sin \bar{\theta_{\mathrm{S}}}\cos [\bar{\phi}(t)-\bar{\phi_{\mathrm{S}}}]}
{2\sin \bar{\theta_{\mathrm{S}}}\sin [\bar{\phi}(t)-\bar{\phi_{\mathrm{S}}}]} \right).
\end{eqnarray}
The Doppler phase $\phi_D =2\pi f R_{AU}\sin \bar{\theta}_{\mathrm{S}}  \cos[\bar{\phi}(t)-\bar{\phi_{\mathrm{S}}}] $. Terms required to define the polarisation angle $\psi_{\mathrm{S}}$ are given as:
\begin{eqnarray}
\hat{\bm{L}}\cdot\hat{\bm{z}}&=&\frac{1}{2}\cos\bar{\theta}_{\mathrm{L}} 
-\frac{\sqrt{3}}{2}\sin\bar{\theta}_{\mathrm{L}} \cos[\bar{\phi}(t)-\bar{\phi}_{\mathrm{L}}], \label{lz} \\
\hat{\bm{L}}\cdot\hat{\bm{N}}&=&\cos\bar{\theta}_{\mathrm{L}}\cos\bar{\theta}_{\mathrm{S}}
+\sin\bar{\theta}_{\mathrm{L}}\sin\bar{\theta}_{\mathrm{S}}\cos(\bar{\phi}_{\mathrm{L}}-\bar{\phi}_{\mathrm{S}}), \label{ln} \\
\hat{\bm{N}}\cdot(\hat{\bm{L}}\times\hat{\bm{z}})&=&\frac{1}{2}\sin\bar{\theta}_{\mathrm{L}}\sin\bar{\theta}_{\mathrm{S}}
\sin(\bar{\phi}_{\mathrm{L}}-\bar{\phi}_{\mathrm{S}}) \notag \\
&&+\frac{\sqrt{3}}{2} \left \lbrace \cos\bar{\theta}_{\mathrm{L}} \sin\bar{\theta}_{\mathrm{S}} \sin [\bar{\phi}(t)- \bar{\phi}_{\mathrm{S}}]
-\cos\bar{\theta}_{\mathrm{S}}\sin\bar{\theta}_{\mathrm{L}} \sin [ \bar{\phi}(t)- \bar{\phi}_{\mathrm{L}}]  \right \rbrace
\end{eqnarray}

\subsubsection*{Geo-centric DECIGO}
In the expressions below $\epsilon \sim 85.6 \degree$ is the angle between the detector plane and ecliptic plane,  $R_E \sim 9000$km is the distance of the detector from the Earth, $R_{\rm AU}$ is the astronomical unit, $\bar{\phi}(t) = 2 \pi t/T$ and $\phi_E = 2 \pi t/T_e$, where $T$ and $T_e$ correspond to 1 year and $\sim$2.36 hours respectively.
\begin{eqnarray}
\cos \theta_{\mathrm{S}}(t)&=&\cos \bar{\theta}_{\mathrm{S}} \left(\frac{1}{2} \cos \epsilon + \frac{\sqrt{3}}{2} \sin \epsilon \cos \phi_E \right) \nonumber \\ &&+ \sin \bar{\theta}_{\mathrm{S}}  \left[\left(-\frac{1}{2} \sin \epsilon + \frac{\sqrt{3}}{2} \cos \epsilon \cos \phi_E \right) \cos[\bar{\phi}(t)-\bar{\phi_{\mathrm{S}}}] \right . \nonumber \\
&&\left .-\frac{\sqrt{3}}{2} \sin \phi_E  \sin[\bar{\phi}(t)-\bar{\phi_{\mathrm{S}}}]\right]  ,  \\
\phi_{\mathrm{S}}(t)&=&\tan^{-1}
\left( \frac{y_s}{x_s} \right),
\end{eqnarray}
where 
\begin{eqnarray}
x_s&=&\cos \bar{\theta}_{\mathrm{S}} \left(-\frac{\sqrt{3}}{2} \cos \epsilon + \frac{1}{2} \sin \epsilon \cos \phi_E \right) \nonumber \\ &&+ \sin \bar{\theta}_{\mathrm{S}}  \left[\left(\frac{\sqrt{3}}{2} \sin \epsilon + \frac{1}{2} \cos \epsilon \cos \phi_E \right) \cos[\bar{\phi}(t)-\bar{\phi_{\mathrm{S}}}] \right . \nonumber \\
&&\left .-\frac{1}{2} \sin \phi_E  \sin[\bar{\phi}(t)-\bar{\phi_{\mathrm{S}}}]\right], \nonumber  \\
y_s&=&-\cos \bar{\theta}_{\mathrm{S}} \sin \epsilon \sin \phi_E  - \sin \bar{\theta}_{\mathrm{S}}  \left[ 
\cos \epsilon \sin \phi_E  \cos[\bar{\phi}(t)-\bar{\phi_{\mathrm{S}}}] \right . \nonumber \\
&&\left .+ \cos \phi_E  \sin[\bar{\phi}(t)-\bar{\phi_{\mathrm{S}}}]\right].
\end{eqnarray}
Terms required to define the polarisation angle $\psi_{\mathrm{S}}$ are given as: 
\begin{eqnarray}
\hat{\bm{L}}\cdot\hat{\bm{z}}&=&\cos \bar{\theta}_{\mathrm{L}} \left(\frac{1}{2} \cos \epsilon + \frac{\sqrt{3}}{2} \sin \epsilon \cos \phi_E \right) \nonumber \\ &&+ \sin \bar{\theta}_{\mathrm{L}} \left[\left(-\frac{1}{2} \sin \epsilon + \frac{\sqrt{3}}{2} \cos \epsilon \cos \phi_E \right) \cos[\bar{\phi}(t)-\bar{\phi_{\mathrm{L}}}] \right . \nonumber \\
&&\left .-\frac{\sqrt{3}}{2} \sin \phi_E  \sin[\bar{\phi}(t)-\bar{\phi_{\mathrm{L}}}]\right] \nonumber \label{lz} \\
\hat{\bm{L}}\cdot\hat{\bm{N}}&=&\cos\bar{\theta}_{\mathrm{L}}\cos\bar{\theta}_{\mathrm{S}}
+\sin\bar{\theta}_{\mathrm{L}}\sin\bar{\theta}_{\mathrm{S}}\cos(\bar{\phi}_{\mathrm{L}}-\bar{\phi}_{\mathrm{S}}), \nonumber  \label{ln} \\
\hat{\bm{N}}\cdot(\hat{\bm{L}}\times\hat{\bm{z}})&=&\sin\bar{\theta}_{\mathrm{L}}\sin\bar{\theta}_{\mathrm{S}}
\sin(\bar{\phi}_{\mathrm{L}}-\bar{\phi}_{\mathrm{S}}) \left(\frac{1}{2} \cos \epsilon + \frac{\sqrt{3}}{2} \sin \epsilon \cos \phi_E \right)   \\
&&+\cos\bar{\phi}(t)\left[\frac{\sqrt{3}}{2} \sin \phi_E (\cos\bar{\theta}_{\mathrm{S}}\cos\bar{\phi}_{\mathrm{L}}\sin \bar{\theta}_{\mathrm{L}} -\cos\bar{\theta}_{\mathrm{L}}\cos\bar{\phi}_{\mathrm{S}}\sin \bar{\theta}_{\mathrm{S}})
\right . \nonumber \\
&&\left . + \left(-\frac{1}{2} \sin \epsilon + \frac{\sqrt{3}}{2} \cos \epsilon \cos \phi_E \right)  (\cos\bar{\theta}_{\mathrm{L}}\sin \bar{\phi}_{\mathrm{S}}\sin \bar{\theta}_{\mathrm{S}} -\sin \bar{\theta}_{\mathrm{L}} \sin \bar{\phi}_{\mathrm{L}}\cos \bar{\theta}_{\mathrm{S}}) \right] \nonumber \\
&&+\sin\bar{\phi}(t)\left[\frac{\sqrt{3}}{2} \sin \phi_E (\cos\bar{\theta}_{\mathrm{S}}\sin\bar{\phi}_{\mathrm{L}}\sin \bar{\theta}_{\mathrm{L}} -\cos\bar{\theta}_{\mathrm{L}}\sin\bar{\phi}_{\mathrm{S}}\sin \bar{\theta}_{\mathrm{S}})
\right . \nonumber \\
&&\left . + \left(-\frac{1}{2} \sin \epsilon + \frac{\sqrt{3}}{2} \cos \epsilon \cos \phi_E \right)  (\sin\bar{\theta}_{\mathrm{L}}\cos \bar{\phi}_{\mathrm{L}}\cos \bar{\theta}_{\mathrm{S}} -\cos \bar{\theta}_{\mathrm{L}} \cos \bar{\phi}_{\mathrm{S}}\sin \bar{\theta}_{\mathrm{S}}) \right] \nonumber \label{nlz},
\end{eqnarray}
and the Doppler phase is given by:
\begin{eqnarray}
\phi_D &=& 2 \pi f \left\lbrace R_E\cos \bar{\theta}_{\mathrm{S}} \sin \epsilon \cos \phi_E \right. \nonumber \\ && \left.+ \sin \bar{\theta}_{\mathrm{S}}  \left[ \left( R_{AU}+ R_E \cos \epsilon \cos \phi_E  \right) \cos[\bar{\phi}(t)-\bar{\phi_{\mathrm{S}}}] \right. \right. \nonumber \\
&&\left .\left. - R_E \sin \phi_E  \sin[\bar{\phi}(t)-\bar{\phi_{\mathrm{S}}}]\right] \right\rbrace.
\end{eqnarray}

\end{document}